\documentclass{emulateapj}

\usepackage{graphics}
\usepackage{epsfig}
\usepackage{natbib}
\shorttitle{High-energy Activity on the 10 Gyr Barnard's Star}
\shortauthors{France et al.}
\begin{document}


\title{The High-Energy Radiation Environment Around a 10 Gyr M Dwarf: Habitable at Last?}


\author{
Kevin France\altaffilmark{1,2,3},
Girish Duvvuri\altaffilmark{2}, 
Hilary Egan\altaffilmark{1},
Tommi Koskinen\altaffilmark{4},
David J. Wilson\altaffilmark{5},
Allison Youngblood\altaffilmark{1},
Cynthia S. Froning\altaffilmark{5},
Alexander Brown\altaffilmark{3},
Juli\'an D. Alvarado-G\'omez\altaffilmark{6,7},
Zachory K. Berta-Thompson\altaffilmark{2,3},
Jeremy J. Drake\altaffilmark{7},
Cecilia Garraffo\altaffilmark{7, 8}, 
Lisa Kaltenegger\altaffilmark{9}, 
Adam F. Kowalski\altaffilmark{1},
Jeffrey L.  Linsky\altaffilmark{10},
R.O. Parke Loyd\altaffilmark{11},
Pablo J. D. Mauas\altaffilmark{12},
Yamila Miguel\altaffilmark{13,14},
J. Sebastian Pineda\altaffilmark{1},
Sarah Rugheimer\altaffilmark{15}, 
P. Christian Schneider\altaffilmark{16},
Feng Tian\altaffilmark{17}, 
Mariela Vieytes\altaffilmark{18}
}


\received{May 9, 2020}
\revised{}
\accepted{August 31, 2020}

\begin{abstract}


Recent work has demonstrated that high levels of X-ray and UV activity on young M dwarfs may drive rapid atmospheric escape on temperate, terrestrial planets orbiting within the habitable zone. However, secondary atmospheres on planets orbiting older, less active M dwarfs may be stable and present more promising candidates for biomarker searches. In order to evaluate the potential habitability of Earth-like planets around old, inactive M dwarfs, we present new {\it Hubble Space Telescope} and {\it Chandra X-ray Observatory} observations of Barnard's Star (GJ 699), a 10 Gyr old M3.5 dwarf, acquired as part of the Mega-MUSCLES program. 
Despite the old age and long rotation period of Barnard's star, we observe two FUV ($\delta_{130}$~$\approx$~5000s; $E_{130}$~$\approx$~10$^{29.5}$ erg each) and one X-ray ($E_{X}$~~$\approx$~10$^{29.2}$ erg) flares, and estimate a high-energy flare duty cycle (defined here as the fraction of the time the star is in a flare state) of $\sim$~25\%. 
A 5 \AA~--~10~$\mu$m spectral energy distribution of GJ 699 is created and used to evaluate the atmospheric stability of a hypothetical, unmagnetized terrestrial planet in the habitable zone ($r_{HZ}$~$\sim$~0.1 AU). Both thermal and non-thermal escape modeling indicate (1) the $quiescent$ stellar XUV flux does not lead to strong atmospheric escape: atmospheric heating rates are comparable to periods of high solar activity on modern Earth, and (2) the $flare$ environment could drive the atmosphere into a hydrodynamic loss regime at the observed flare duty cycle: sustained exposure to the flare environment of GJ 699 results in the loss of $\approx$~87 Earth atmospheres Gyr$^{-1}$ through thermal processes and $\approx$~3 Earth atmospheres Gyr$^{-1}$ through ion loss processes, respectively. These results suggest that if rocky planet atmospheres can survive the initial $\sim$~5 Gyr of high stellar activity, or if a second generation atmosphere can be formed or acquired, the flare duty cycle may be the controlling stellar parameter for the stability of Earth-like atmospheres around old M stars. 



\end{abstract}

\clearpage

\section{Introduction}

The nearest known terrestrial planets in the liquid water habitable zone (HZ) orbit M dwarf stars (T$_{eff}$ $\leq$ 4000 K), e.g., Proxima Cen b~\citep{anglada16} and the TRAPPIST-1 planets~\citep{gillon17}.  The {\it Transiting Exoplanet Survey Satellite} ($TESS$) is predicted to detect an additional $\sim$~15 planets in the HZ of M dwarfs in the near future (Barclay et al. 2018, Huang et al. 2018). Due to more detectable atmospheric features with transit spectroscopy observations than Sun-Earth analogs (owing to the smaller stellar radii), rocky planets around M dwarfs will likely be the only potentially habitable planets whose atmospheres could be successfully searched for signs of life in the near future (with the {\it James  Webb Space Telescope} ($JWST$) and Extremely Large Telescopes  (ELTs) on the ground; Snellen et al. 2015; Barstow \& Irwin 2016; Morley et al. 2017), prior to a large ultraviolet/optical/infrared mission in the 2030s or 2040s (LUVOIR STDT Final Report).\nocite{anglada16, gillon17, sullivan15, barclay18, barstow16, morley17} 

The long-term stability of Earth-like atmospheres depends critically on the X-ray (1~--~100~\AA) and EUV (100~--~911~\AA) irradiance~\citep{tian08,johnstone15}. Elevated EUV flux from the young Sun~\citep{tu15} could have led to 10 times greater oxygen loss rates and 90 times greater carbon loss rates on the early Earth by increasing the suprathermal population of these atoms~\citep{Amerstorfer17}. In highly irradiated planets, the outflow can be sufficiently rapid that heavier elements (O and C) are dragged along through collisions with the lighter hydrogen atoms, as observed on hot Jupiters~\citep{vidal03,linsky10,ballester15}. Free electrons produced by stellar EUV photons can attain altitudes much greater than ions, producing an ambipolar electric field that leads to a non-thermal ionospheric outflow (O$^{+}$ and N$^{+}$ winds; Kulikov et al. 2006; Lichtenegger et al. 2016; Dong et al. 2017).\nocite{kulikov06,lichtenegger16,dong17} XUV ($\equiv$ X-ray + EUV) irradiance in the HZ environments around M dwarfs is expected to be more than a factor of 10 times the average Earth-Sun value owing to the close-in habitable zone~\citep{france16}, making M dwarf exoplanets particularly prone to XUV-driven atmospheric escape. Additionally, the XUV luminosity of M dwarfs is enhanced by another factor of 10 - 50 relative to solar-type stars during their prolonged pre-main-sequence evolution~\citep{shkolnik14,luger15,ribas17,peacock20}.  The high-energy emission stays in the saturated regime of rotation-driven activity on M dwarf systems several Gyrs longer than that of Sun-like stars (see Pineda et al. 2020 and references therein), well into the era in which life emerged on Earth.\nocite{pineda20}

The discovery of rocky planets in the HZs of nearby M dwarfs has motivated new atmospheric mass loss calculations that highlight the need for empirically-based XUV irradiance spectra to estimate their long-term habitability. For instance, volatile loss on Proxima Centauri b, the nearest potentially habitable planet, could be catastrophic or small depending on assumptions about the planetary initial conditions, atmospheric composition, and the poorly constrained pre-main-sequence evolution of M dwarfs (e.g., Ribas et al. 2016a). 

\citet{garcia17} presented a study of EUV-driven proton and O$^{+}$ escape from an Earth-like planet orbiting Proxima Cen. They found mass loss rates several orders of magnitude higher than present-day Earth for EUV fluxes 10 - 20 times the current day EUV solar irradiance. This escape results in complete loss of a one bar atmosphere in less than 0.5 Gyr, even in the presence of a terrestrial-strength magnetic field. \citet{airapetian17} reached similar conclusions on the loss of one bar of H and O via ion escape. Their models found that basal EUV fluxes of M dwarfs removed a complete Earth-like atmosphere in a few hundred Myr. They argued that elevated EUV fluxes from persistent flares or the long pre-main-sequence phase of M dwarfs could effectively render all Earth-like planets around M dwarfs barren in the absence of internal or external resupply of volatiles. 

With several mechanisms demonstrated to be efficient at driving rapid atmospheric escape from HZ planets around M dwarfs, it is worth considering under what conditions atmospheres $could$ be stable on these planets. This is particularly relevant now as the astronomical community evaluates the most promising targets in which to invest the limited observing resources of $JWST$. In this paper, we investigate an ``old'' M star and the effect of this star on a hypothetical planet located in its habitable zone.  We ignore the likely extreme early history of the planet to ask if a secondary atmosphere could be retained later in life, 5~--~10~Gyr after formation. Significant second generation atmospheres may be produced by outgassing of material accreted as solids in impacts or outgassing through volcanic processes~\citep{holland02,moore20}. Therefore, the present-day conditions experienced by M dwarf planets are likely to be equally or more important to atmospheric detectability than the extreme conditions experienced earlier in their evolution. 

Towards this end, we present new X-ray and UV observations of Barnard's Star, an old M3.5 dwarf star (7~--~12 Gyr; Benedict et al. 1998; Gauza et al. 2015).  Mid-M dwarfs are near the peak of the red dwarf mass function~\citep{bochanski10}, with M dwarfs themselves making up the majority of stars in the universe.  The main sequence lifetimes of these stars are longer than the age of the Galaxy and exoplanet detection surveys typically target older stars to mitigate contamination from stellar activity. Combining these considerations with $Kepler$ results demonstrating a relatively high fraction of terrestrial planet occurrence~\citep{dressing15}, old mid-M dwarfs are important targets for rocky planet detection and characterization efforts.    
   
We use these new stellar observations of Barnard's Star to calculate the thermal and non-thermal atmospheric mass loss rates from an Earth-like planet in the HZ of Barnard's Star. This work is meant to explore the stellar habitability of rocky planets in the habitable zones around old M dwarfs in general. For the connection between the stellar properties of GJ 699 with its known planet, GJ 699 b, we refer the reader to the upcoming work of Guinan and collaborators. The present work is laid out as follows: we present a brief overview of the Barnard's Star in Section 2. The new $HST$ and $Chandra$ observations are presented in Section 3, and details of the spectral energy distribution and high-energy variability of Barnard's star are presented in Section 4. The stability and evolution of Earth-like planets in this environment are presented in Section 5 and Section 6 gives a brief summary of this work.

\begin{figure*}[t]
\begin{center}
\hspace{-0.3in}\epsfig{figure=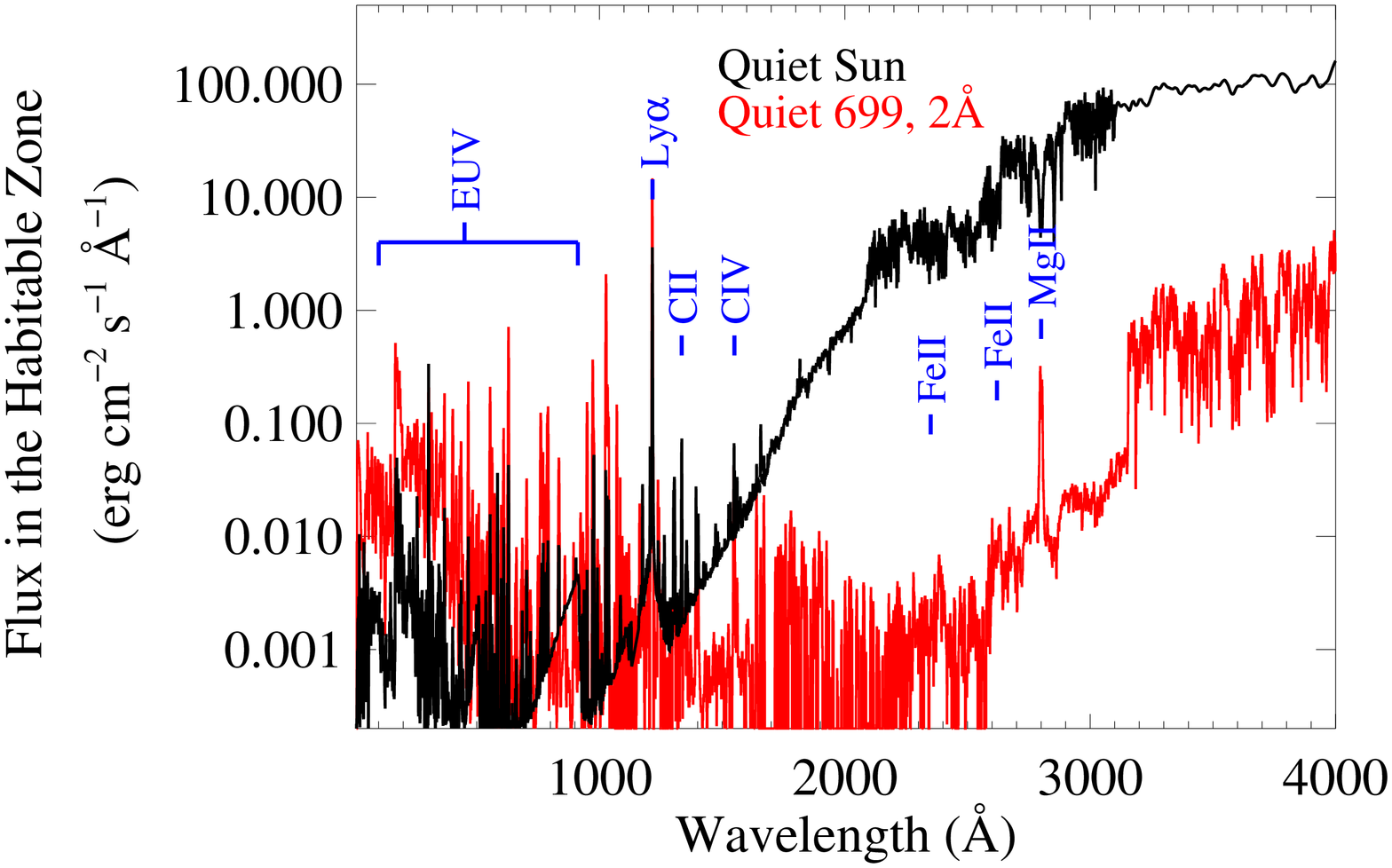,width=2.6in,angle=90}
\epsfig{figure=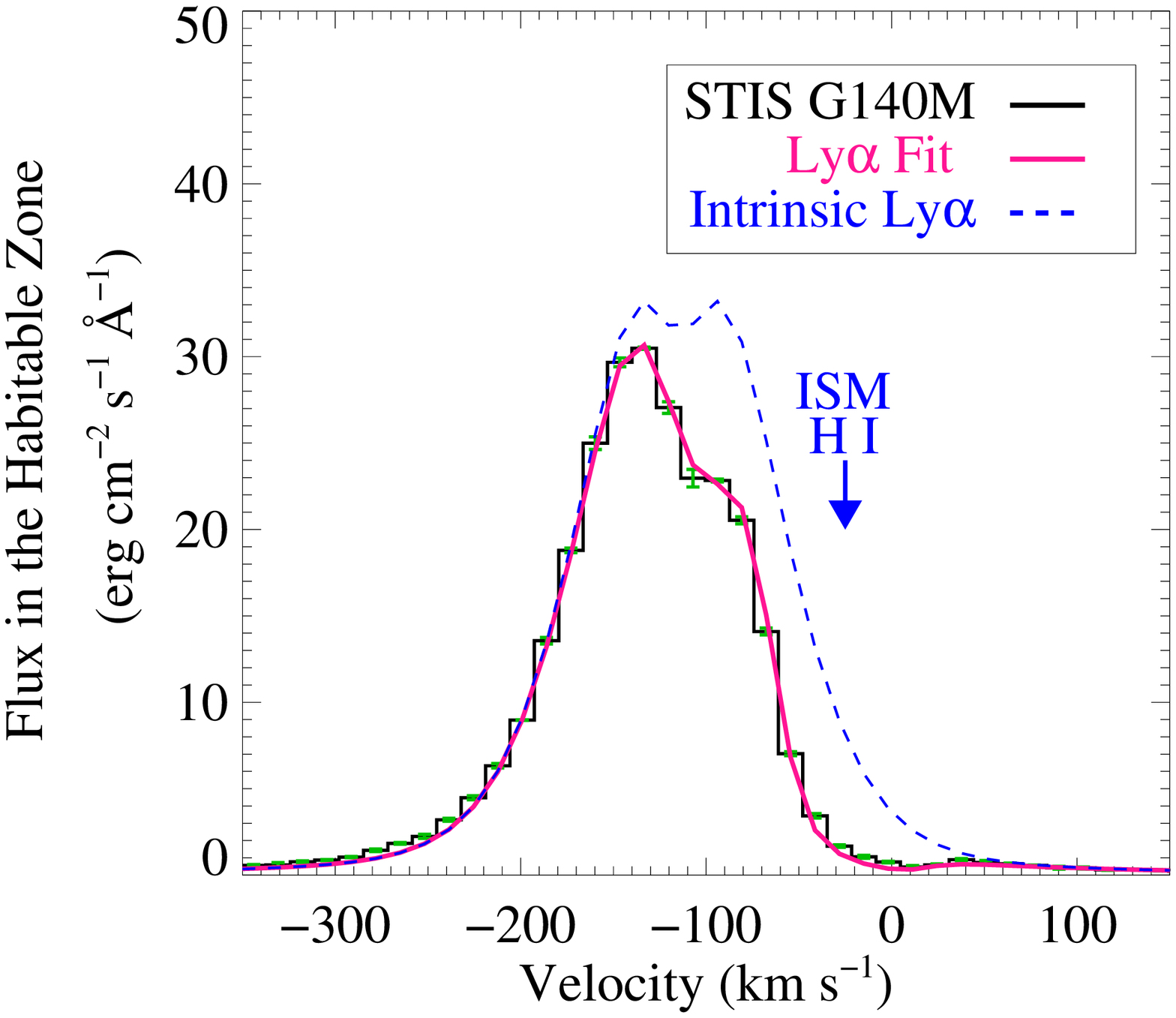,width=2.6in,angle=90}
\vspace{+0.0in}
\caption{
\label{cosovly} 
($left$) A comparison between the ultraviolet spectrum of the quiet Sun (black, from Woods et al. 2009) and the quiescent flux from Barnard's Star (GJ 699, shown in red binned to the 2~\AA\ resolution of the solar data). Prominent emission lines are labeled. The reconstructed EUV flux (Section 4.3) is enhanced relative to the quiet Sun, whereas the FUV flux (mainly chromospheric and transition region emission lines; Section 4) is comparable and the NUV flux is below the Sun by 2~--~3 orders of magnitude owing to the cooler effective temperature of the M3 star.
($right$) Ly$\alpha$ is the brightest line in the UV spectrum of M dwarfs, we show the observed (black histogram), reconstructed (blue dashed line; Section 4), and model fit (pink solid line) Ly$\alpha$ emission line, employing the reconstruction technique of~\citet{youngblood16}. The reconstruction parameters find an intrinsic line flux, $F$(Ly$\alpha$)~=~1.02~$\times$~10$^{-12}$ erg cm$^{-2}$ s$^{-1}$ with an interstellar column density of log$_{10}$$N$(HI)~=~17.60.  Note that the large negative radial velocity of GJ 699 moves the star mostly out of the interstellar \ion{H}{1} line core.   
}
\end{center}
\end{figure*}

\section{Barnard's Star~--~Prototype of Mature Star-Planet Environment}

Barnard's Star, GJ 699, is an intermediate mass M dwarf (M3.5V) with approximately 16\% of the mass of the Sun~\citep{ribas18}. It is particularly notable for its proximity to Earth ($d$~=~1.83 pc; Gaia collaboration) and old age. Previous work has estimated the age of GJ 699 to be between 7~--~12 Gyr, based on a combination of slow rotation period ($P_{rot}$ = 130~--~145 days; Benedict et al. 1998; Toledo-Padron et al. 2019), low X-ray luminosity~\citep{stelzer13,guinan19}, and low-levels of magnetic activity~\citep{hunsch99,ribas18}.  Despite the low activity levels, long-term optical monitoring observations have recorded Balmer and metal line flaring from this star~\citep{paulson06}.  For the purposes of this paper, we assume that GJ 699 is roughly twice the age of the Sun and is therefore a prototypical object for measuring the high-energy properties of old, mid-M dwarf stars. The bolometric luminosity of GJ 699 is 1.2~--~1.3~$\times$~10$^{31}$ erg s$^{-1}$ (approximately 0.0033 $L_{\odot}$; Ribas et al. 2018).\nocite{toledo19} 

High-precision radial velocity monitoring has recently shown that GJ 699 hosts a cold, super-Earth mass ($M$ {\it sin i} = 3.23~$\pm$~0.44 M$_{\oplus}$) planet, GJ 699 b~\citep{ribas18}. GJ 699 b orbits with a 233 day period, which suggests an equilibrium temperature of $\sim$~100 K given the cool photospheric temperature of GJ 699 (3100~--~3300 K; Dawson \& De Robertis 2004; Toledo-Padron et al. 2019; Ribas et al. 2018).

\section{$HST$ and $Chandra$ Observations}

GJ 699 was observed between March and June 2019 with the {\it Hubble Space Telescope} and the {\it Chandra X-ray Observatory} as part of the Mega-MUSCLES\footnote{Mega-Measurements of the Ultraviolet Spectral Characteristics of Low-mass Exoplanetary Systems} observing program (see Froning et al. 2019)\nocite{froning19}. The observations (Table 1) were designed to record the highest fidelity UV and X-ray observations of GJ 699 obtained to date, and extend existing panchromatic surveys of M dwarfs~\citep{france16}. In order to obtain a full census of the UV emission incident in the environment of GJ 699, Mega-MUSCLES uses $HST$ to obtain spectral coverage from 1100~--~5500~\AA: the G130M and G230L modes of COS, and the G140M, G140L, G230L and G430L modes of STIS provide spectral coverage across this bandpass~\citep{loyd16}. We use the G140M mode of STIS with the 52\arcsec~$\times$~0.1\arcsec\ slit to measure the Ly$\alpha$ profile in order to minimize the contribution of geocoronal
emission.  Resonant scattering of Ly$\alpha$ in the local ISM requires that the line must be reconstructed to provide a reliable measure of the intrinsic Ly$\alpha$ radiation field for exoplanet atmosphere calculations~\citep{youngblood16}. We obtained 5932 s of STIS G140M observations on 04 March 2019. 

In the FUV (except Ly$\alpha$) we use COS G130M and STIS G140L observations to cover 1100~--~1700~\AA\ and monitor the target for flares. We obtained 12902 s of COS G130M and 7018 s of STIS G140L observations on 04 March 2019. The high-sensitivity COS G130M mode~\citep{green12} allows us to track temporal variability in 10$^{4-7}$ K chromospheric, transition region, and coronal activity indicators (using the \ion{C}{2}, \ion{Si}{3}, \ion{C}{3}, \ion{N}{5}, \ion{Fe}{19}, and \ion{Fe}{21} emission lines) over an 8-hour interval (5 contiguous spacecraft orbits). This observing method has been used to study the flare properties of a range of nearby M dwarfs~\citep{loyd18a,froning19}. 

At NUV wavelengths, we use STIS G230L ($\lambda$~$>$~2200~\AA) to observe the NUV continuum, \ion{Fe}{2} $\lambda$2400 and $\lambda$2600, and \ion{Mg}{2} $\lambda$2800, but take advantage of the superior sensitivity of the COS G230L mode to observe the 1750~--~2200~\AA\ region. We obtained 200 s of STIS G230L, 318 s of COS G230L, and 5 s of STIS G430L observations on 04 March 2019. 

GJ 699 was observed with the $Chandra$ ACIS-S back-illuminated S3 chip on 17 June 2019. The observed X-ray and FUV observations are combined to constrain differential emission measure (DEM) calculations (see Section 4.2 and Duvvuri et al. - in prep). The X-ray observations of the Mega-MUSCLES sources will be analyzed in the context of chromospheric and coronal evolution in a future work (Linsky et al. - in prep. and Brown et al. - in prep.).


\section{The UV and X-ray Radiation Environment}  

The observed UV and X-ray spectra of GJ 699 are qualitatively similar to other optically inactive M dwarfs.  However, the average UV emission line luminosities (combining flare and quiescent periods to be consistent with previous work) of GJ 699 are among the lowest ever measured for an M dwarf~\citep{youngblood17, melbourne20}. 
The luminosity of well-studied FUV ionic emission lines are, e.g., \ion{N}{5} (log$_{10}$(L(\ion{N}{5})) = 23.9), \ion{C}{2} (log$_{10}$(L(\ion{C}{2})) = 24.1), \ion{C}{4} (log$_{10}$(L(\ion{C}{4})) = 24.5), and \ion{Mg}{2} (log$_{10}$(L(\ion{Mg}{2})) = 25.7).  Interstellar attenuation by low-ionization metals is not significant owing to the high radial velocity of GJ 699.  The full list of observed UV emission line fluxes and X-ray data are given in Table 2. We ascribe the low UV emission line luminosity to decreasing activity with age~\citep{guinan16}. The proximity of GJ 699 allows us to observe the star at comparable signal-to-noise despite the lower absolute flux levels. 

Figure 1 shows the full UV (100~$<$~$\lambda$~$<$~4000~\AA) spectral energy distribution of GJ 699 and the reconstructed Ly$\alpha$ profile. The GJ 699 spectrum is compared to the UV spectrum of the quiet Sun~\citep{woods09}, scaled to the 1 AU equivalent distance. The comparison with the Sun is typical for relatively inactive M dwarfs~\citep{france12a,france16}; the reconstructed EUV flux is enhanced relative to the quiet Sun (Section 4.3), whereas the FUV flux (mainly chromospheric and transition region emission lines) is comparable and the NUV flux is below the Sun by 2~--~3 orders of magnitude owing to the cooler photosphere of the M3 star. The Ly$\alpha$ reconstruction is based on the framework presented by~\citet{youngblood16}, where the intrinsic Ly$\alpha$ flux is determined by simultaneously fitting interstellar hydrogen and deuterium absorption and the intrinsic emission line shape. The deuterium fraction is fixed at D/H~=~1.5~$\times$~10$^{-5}$~\citep{wood04} and does not contribute appreciably to the line shape for the low \ion{H}{1} column density and resolution of the STIS G140M data.  The Ly$\alpha$ fit parameters are $F$(Ly$\alpha$)~=~1.02 ($\pm$0.01)~$\times$~10$^{-12}$ erg cm$^{-2}$ s$^{-1}$, log$_{10}$$N$(HI)~=~17.60~$\pm$~0.02, Doppler $b$-value $b_{HI}$~=~8.6$^{+0.4}_{-0.5}$ km s$^{-1}$, and the velocity of the interstellar \ion{H}{1} absorption is $v_{HI, ISM}$~=~$-$21.0~$\pm$~1.1 km s$^{-1}$. 

The low X-ray luminosity of GJ 699 has been one of the factors used to argue for an old age for this star (e.g., Ribas et al. 2018 and references therein). Our new $Chandra$ observations confirm weak, but non-zero quiescent X-ray flux from this source (0.3~--~10 keV flux, $F_{X}$~$\approx$~4.8~$\times$~10$^{-14}$ erg cm$^{-2}$ s$^{-1}$; log$_{10}$($L_{X}$)~=~25.3; $L_{X}$/$L_{bol}$ = 1.6~$\times$~10$^{-6}$). The X-ray luminosity we measure with $Chandra$ is within a factor of two compared to previous $ROSAT$ data (log$_{10}$($L_{X}$)~=~25.6); we note that X-ray luminosity measurements below 10$^{26}$ erg s$^{-1}$ are amongst the lowest in existing large samples of M dwarfs~\citep{pizzolato03,stelzer13}, even for slowly rotating stars. 


\begin{figure}
\begin{center}
\epsfig{figure=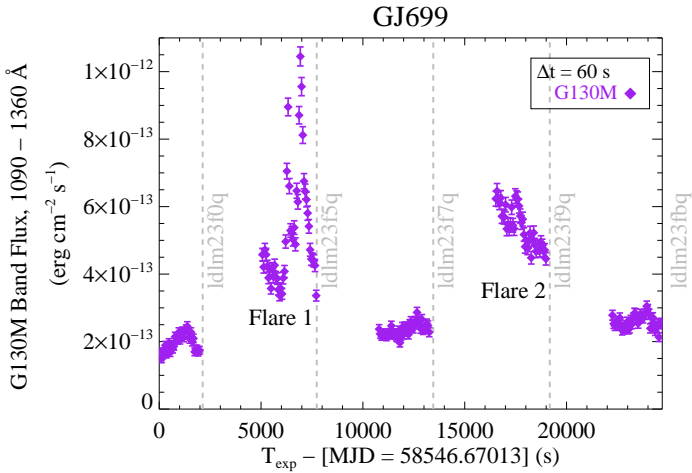,width=2.55in,angle=90}
\vspace{+0.0in}
\caption{
\label{cosovly} 
$HST$-COS G130M, 1090~--~1360~\AA\ band-integrated lightcurve of Barnard's Star (GJ~699) at 60 s cadence, with individual data sets labeled in gray. Gaps in coverage wavelengths between approximately 1209~--~1225~\AA\ fall in the COS detector ``gap'' and therefore these data do not include flux from the bright \ion{H}{1} Ly$\alpha$ emission line. Two prominent flares are observed in this $\sim$~25 ksec monitoring visit: ``Flare 1'' is dominated by chromospheric line variability while ``Flare 2'' shows strong transition region and coronal line enhancement (Figure 3).  
}
\end{center}
\end{figure}


\begin{figure*}
\begin{center} 
\epsfig{figure=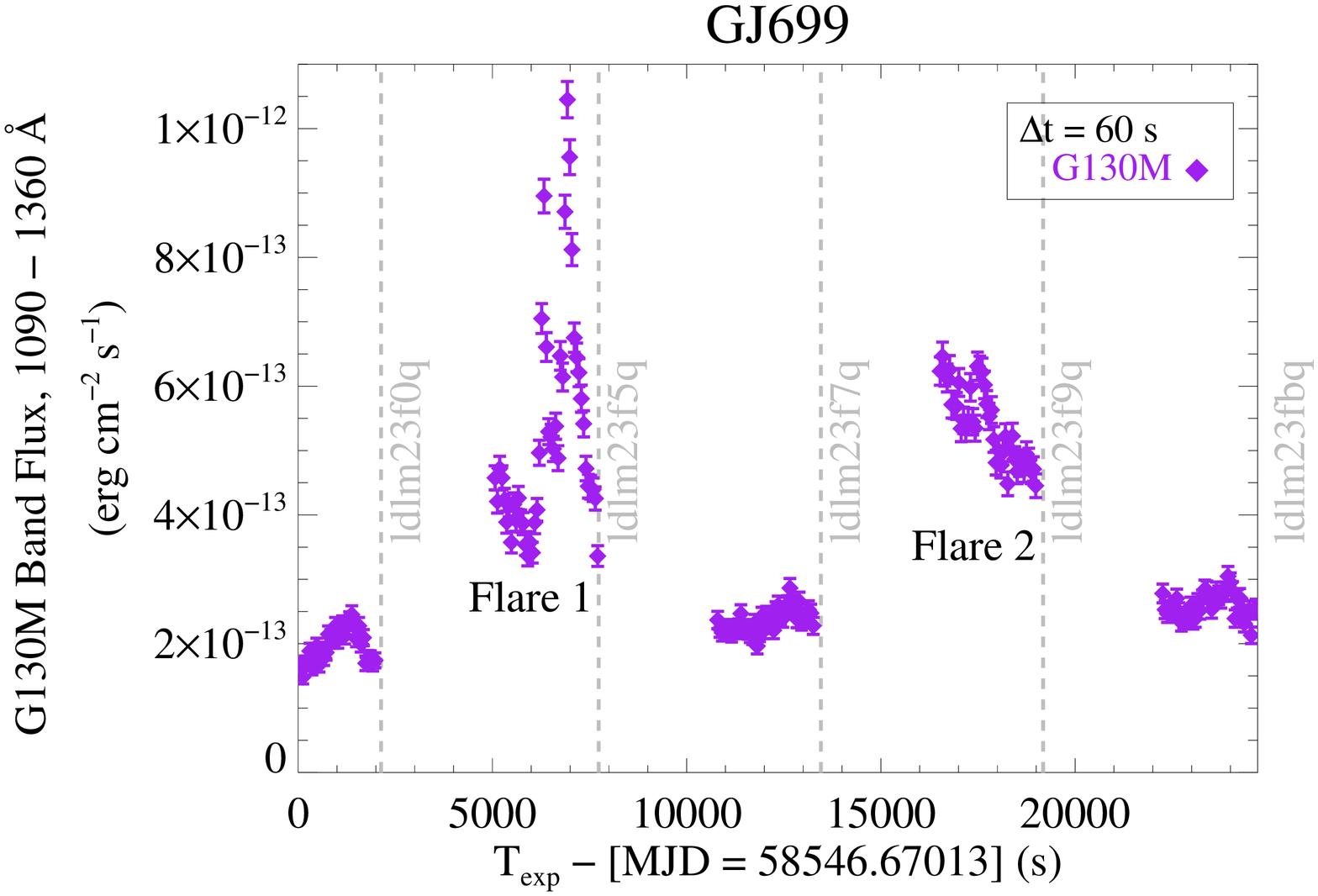,width=2.25in,angle=90} \hspace{0.0in}
\epsfig{figure=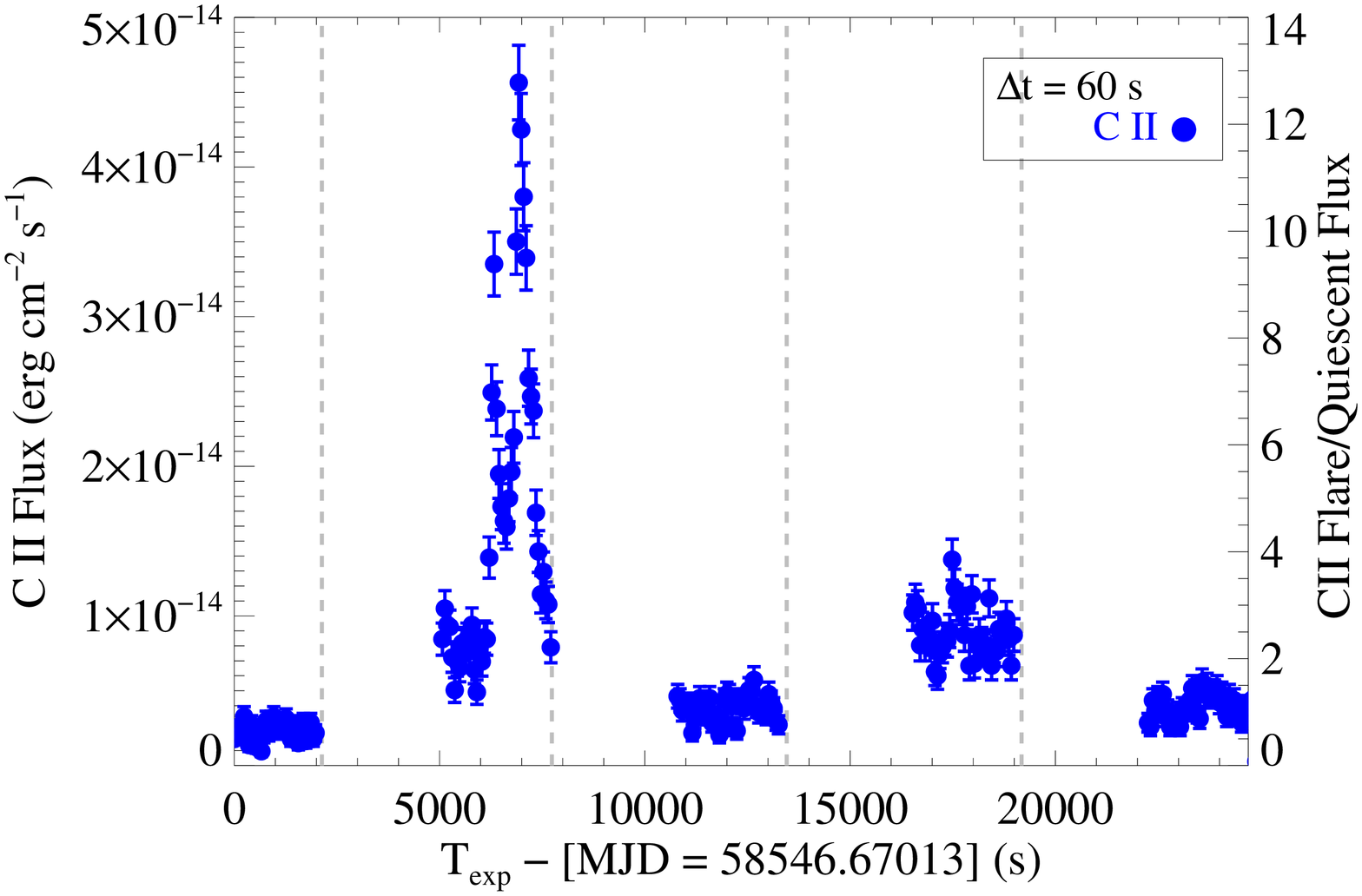,width=2.25in,angle=90}
\epsfig{figure=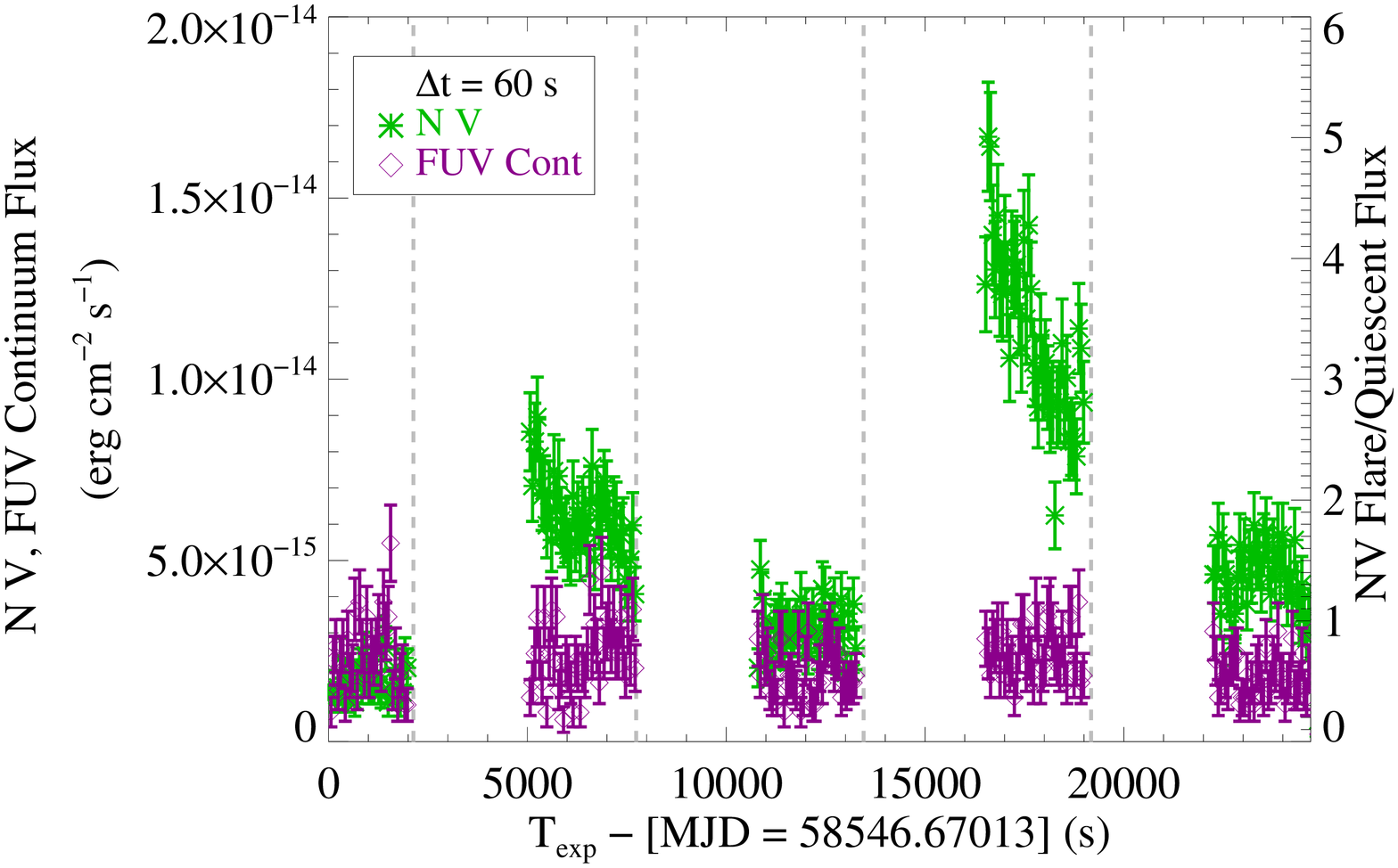,width=2.25in,angle=90} \hspace{-0.0in}
\epsfig{figure=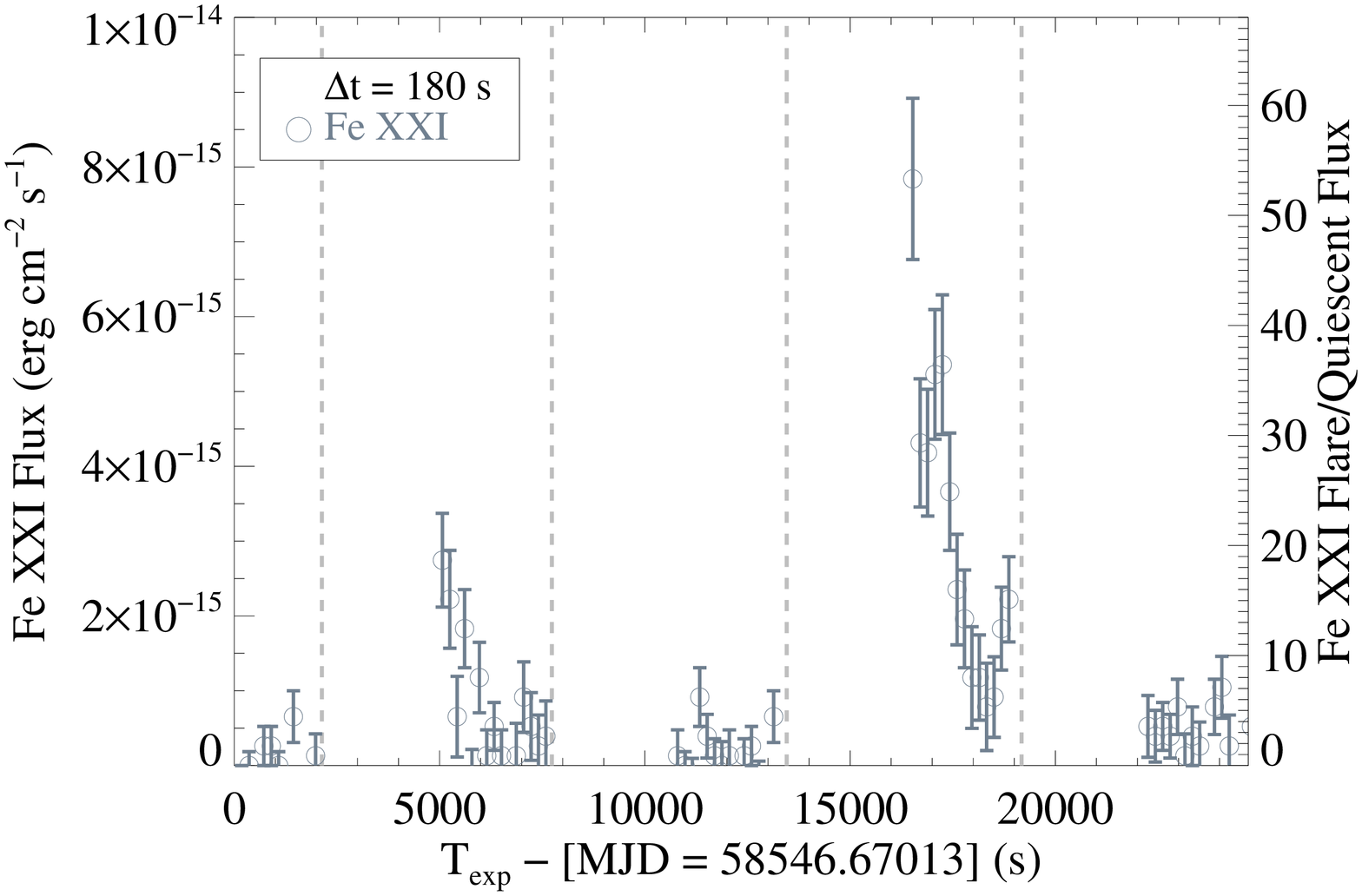,width=2.25in,angle=90}
\vspace{+0.0in}
\caption{
\label{cosovly} 60 s cadence simultaneous $HST$-COS lightcurves from the $\sim$~25 ksec monitoring visit on GJ 699, with different spectral features isolated. The flux levels are shown on the left axes and the quiescent-normalized values are shown on the right axes.  ({\it upper left}): G130M, 1090~--~1360~\AA\ band-integrated lightcurve, reproduced from Figure 2 for context. ({\it upper right}): chromospheric \ion{C}{2} $\lambda$1335\AA\ multiplet, in part driving the overall response of Flare 1. ({\it lower left}): transition region \ion{N}{5} $\lambda$1240\AA\ doublet (green) and the FUV continuum from approximately 1088~--~1108~\AA\ (dark purple). These data show a modest (2~$\times$) brightness enhancement in \ion{N}{5} in Flare 1 with a stronger (6~$\times$) brightness enhancement in Flare 2 that follows the shape of the coronal emission and drives the overall response of Flare 2. The FUV continuum does not show variability with either the chromospheric or coronal components, in contrast to recent observations of GJ 674~\citep{froning19} and J02365171-5203036~\citep{loyd18b}. ({\it lower right}): lightcurve of coronal \ion{Fe}{21} $\lambda$1354~\AA\ emission (sampled at 180s cadence due to lower flux levels), qualitatively similar to that of \ion{N}{5}. 
}
\end{center}
\end{figure*}

\subsection{Active FUV Flaring on Barnard's Star}

Far-UV lightcurves of GJ 699 were created in individual stellar emission lines as well as the band-integrated ``G130M Flux''~\citep{loyd18a} for the 5-orbit flare monitoring observations (Section 3). The G130M Flux (1090~--~1360~\AA; excluding Ly$\alpha$) light curves are shown in Figure 2, with the end of each orbit noted by a gray dashed line and associated observation identifier. Orbits 1, 3, and 5 show relatively constant emission while obvious flares occurred in orbits 2 (hereafter ``Flare 1'') and 4 (hereafter ``Flare 2''). Approximately 40\% of the total exposure time of the FUV monitoring observations are strongly influenced by flares, this fraction being comparable to the FUV flare rates observed by~\citet{loyd18a} using a similar observing cadence. We describe the UV flaring in Section 4.1 and the X-ray flaring in Section 4.2.  The general  result is that UV and X-ray flares on Barnard's star are comparable to those observed on younger M-dwarfs, albeit at lower absolute intensity owing to the lower basal flux level.  Our new observations of GJ 699 indicate that FUV and X-ray flaring rates on M dwarfs are maintained out to ages of $\sim$~10 Gyr, and continue to be significantly greater than the Sun (Section 4.2). 

One notices in Figure 2 that the temporal shapes of Flare 1 and Flare 2 are markedly different. We created lightcurves in individual spectral lines to investigate this phenomena, observing that Flare 1 is dominated by chromospheric ``low ions'' (\ion{C}{2}, \ion{C}{3}, \ion{Si}{3}; Figure 3, upper right), while Flare 2 is dominated by transition region and coronal line emission (\ion{N}{5}, \ion{Fe}{21}; Figure 3, bottom row).  Quiescent-normalized lightcurve values are shown on the right axes in Figure 3.   

 Despite their different spectral character, the quiescence-subtracted, G130M band equivalent flare durations ($\delta_{130}$~$\approx$~5000s; see Hawley et al. 2014 for a description of the flare duration) and integrated G130M-band energies for Flare 1 and Flare 2 are nearly identical (log$_{10}$ $E_{130}$~=~29.54 and 29.53). 
Unlike the very strong coronal flare recently observed on GJ 674~\citep{froning19} and the high-luminosity FUV flare observed on the young M star J02365171-5203036~\citep{loyd18b}, the GJ 699 flares were not accompanied by a large increase in the FUV continuum emission.\nocite{hawley14}

\begin{figure*}
\begin{center}
\epsfig{figure=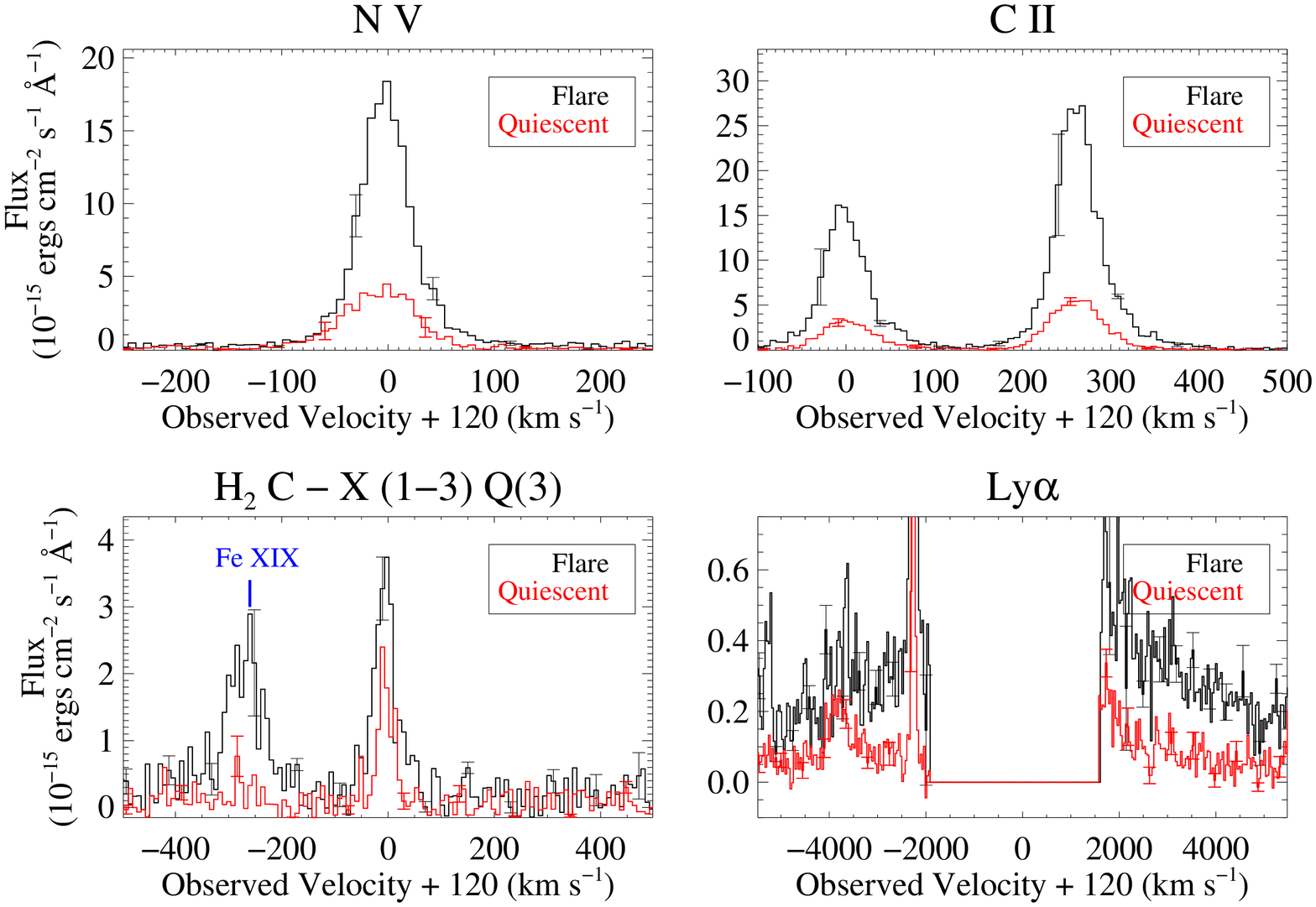,width=5.0in,angle=90}
\vspace{-0.1in}
\caption{
\label{cosovly} Flare (black) and quiescent (red) line profiles of prominent emission lines with different formation temperatures are plotted as a function of velocity. Velocity scales are shifted to the rest frame of the star, $v_{GJ 699}$~=~$-$120 km s$^{-1}$. ({\it upper left}): \ion{N}{5} $\lambda$1238.82\AA, log$_{10}$$T_{form}$~$\approx$~5.2, formed in the transition region and lower corona. ({\it upper right}): \ion{C}{2} multiplet (centered on $\lambda$1334.53\AA), log$_{10}$$T_{form}$~$\approx$~4.5, formed in the chromosphere and transition region. ({\it lower left}): H$_{2}$ $C$~--~$X$~(1~--~3) Q(3) $\lambda$1119.08\AA, log$_{10}$$T_{form}$~$\approx$~3.0, likely formed in a temperature minimum in the lower chromosphere. ({\it lower right}): the broad wings of \ion{H}{1} Ly$\alpha$ $\lambda$1215.67\AA, log$_{10}$$T_{form}$~$\approx$~4~--~5, formed in several regions of the stellar atmosphere. The Ly$\alpha$ line core was not covered by our $HST$-COS observations. }
\end{center}
\end{figure*}

\begin{figure*}
\begin{center}
\epsfig{figure=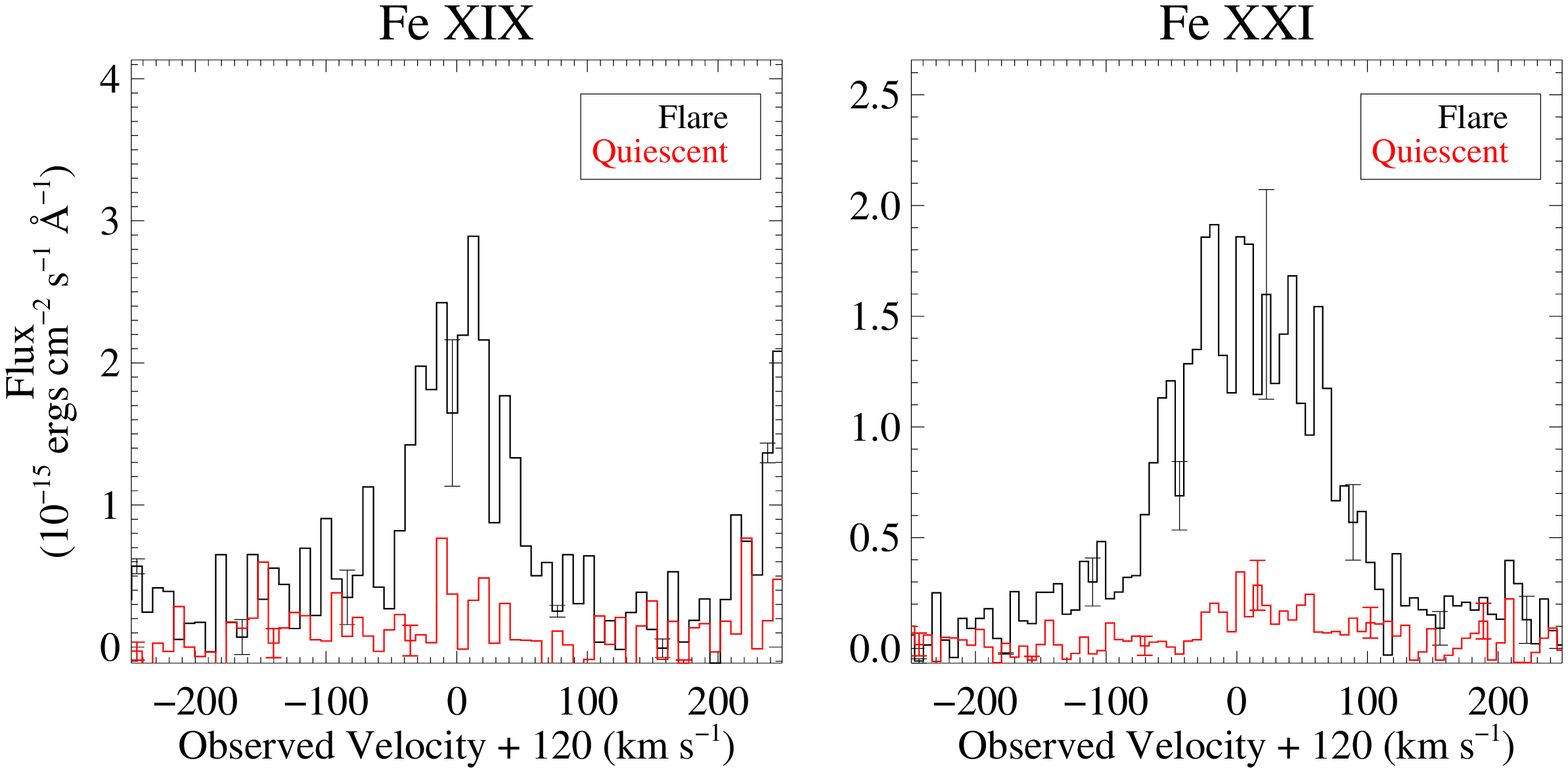,width=5.0in,angle=90}
\vspace{-0.75in}
\caption{
\label{cosovly} Flare (black) and quiescent (red) line profiles of coronal iron lines ($left$: \ion{Fe}{19}, $\lambda$1118.06~\AA; $left$: \ion{Fe}{21}, $\lambda$1354.08~\AA) as a function of velocity. Velocity scales are shifted to the rest frame of the star, $v_{GJ 699}$~=~$-$120 km s$^{-1}$.}
\end{center}
\end{figure*}

To compare the average flare and quiescent line profiles of GJ 699, we have combined Flare 1 and Flare 2 observations into a master flare spectrum, to be compared with the master quiescent spectrum (a coaddition of orbits 1, 3, and 5). Figure 4 demonstrates the line-profile changes between the flare and quiescent states of GJ 699 for atomic species with a range of ionization states. 
The average \ion{N}{5} doublet flux increases by a factor of 3.5 in flare periods (Table 2).  \citet{wood97} has shown that a two-Gaussian fit is able to capture the full range of line broadening seen in the transition region spectra of many cool stars.  While the data quality was not high enough to reliably fit two components to the quiescent spectrum of GJ 699, the flare state appears to be associated with an increase in a narrow component to the line profile which reduces the average FWHM of the Gaussian fit from FWHM$_{quiesc}$(N V)~$\approx$~67~$\pm$~5 km s$^{-1}$ to FWHM$_{flare}$(N V)~$\approx$~47~$\pm$~2 km s$^{-1}$. The average \ion{C}{2} multiplet flux increased by an average of 4.2 in the flare state, with a clear two-component structure developing in the flare spectrum. The FWHM$_{quiesc}$(C II)~$\approx$~59~$\pm$~3 km s$^{-1}$ evolves to a clear narrow FWHM$_{flare,narrow}$(C II)~$\approx$~41~$\pm$~3 km s$^{-1}$ and broad FWHM$_{flare,broad}$(C II)~$\approx$~95~$\pm$~10 km s$^{-1}$ profile, with the broad profiles being redshifted by 9 and 19 km s$^{-1}$ relative to the narrow components for the $\lambda$1334.53 and $\lambda$1335.71 lines, respectively. \nocite{wood97} The \ion{C}{2} flare spectra also show a red wing enhancement observed in other M dwarf flares that is attributed to ``chromospheric condensation'' as the electron beam from the flare heats and compresses material at the footpoints, pushing it toward the star~\citep{hawley03}.
Finally, the flare spectrum also shows a notable enhancement in the very broad Ly$\alpha$ wings, $|$$v_{Ly\alpha}$$|$~$>$ 4000 km s$^{-1}$ (Figure 4, lower right), discussed by~\citet{youngblood16}.  A detailed study of the Ly$\alpha$ line broadening mechanism is beyond the scope of this work, but we note the similarities to the Stark broadened Balmer lines observed on GJ 699 by~\citet{paulson06}.

Werner band H$_{2}$ fluorescence lines $C$~--~$X$ (1~--~3) Q(3) (1119~\AA) and $C$~--~$X$ (1~--~4) Q(3) (1163~\AA) are clearly detected in both quiescent and flare states~\citep{redfield02b,france07}, with factors of 1.5~--~2 flux increase during the flare. These lines are pumped by the \ion{O}{6} 1032~\AA\ emission line and are likely formed in the temperature minimum in the lower stellar chromosphere or in starspots~\citep{kruczek17}. We note that the \ion{O}{6}-pumped H$_{2}$ features that originate in $v^{`'}$~=~1 are stronger than the Ly$\alpha$-pumped features that originate in $v^{`'}$~=~2, different than in a typical low-activity M dwarf but similar to the behavior seen in AU Mic~\citep{france07}. Assuming that the intrinsic Ly$\alpha$ spectrum dominates the total stellar FUV emission (e.g., France et al. 2012)\nocite{france12a}, we interpret the observation of fluorescent H$_{2}$ originating from the lower-excitation $v^{`'}$~=~1 state and the absence of fluorescent H$_{2}$ originating from the higher-excitation $v^{`'}$~=~2 as evidence for a low-temperature ($T$(H$_{2}$)~$\leq$~1500 K) molecular component to the atmosphere of Barnard's star.  
The physical location within the stellar atmosphere of this cool component is not clear and may provide interesting constraints on future models of M dwarf atmospheres~\citep{fontenla16,peacock19}.  While 1500~K is cooler than the 2600 K chromospheric temperature minimum estimated from the M1.5V model atmosphere of~\citet{fontenla16}, it is consistent with the chromospheric minimum found in the M8V model atmosphere calculated by~\citet{peacock19}.

High temperature ($T_{form}$~=~10$^{6-7}$ K) iron ionization states in the $HST$-COS band, \ion{Fe}{12} $\lambda$$\lambda$1242, 1349, \ion{Fe}{19} $\lambda$1118, and \ion{Fe}{21} $\lambda$1354, are observed in some active cool stars~\citep{redfield02b, ayres10, froning19}, including the Sun during flares (e.g., Battaglia et al. 2015).\nocite{battaglia15} The observation of strong, coronal \ion{Fe}{21} emission in the time-averaged FUV spectra of this old, relatively inactive star was the first indication that flare activity had been present during our FUV observations. The \ion{Fe}{12}, \ion{Fe}{19}, and \ion{Fe}{21} fluxes observed in the average flare spectra are 1.4, 6.2, and 9.4 times higher, respectively, than the upper limits measured in the average quiescent data. Figure 5 illustrates the flare versus quiescent behavior of the \ion{Fe}{19} and \ion{Fe}{21} lines observed in the $HST$-COS data. The flare/quiescent flux ratio of \ion{Fe}{21} is uncertain owing to the very low signal levels in the quiescent state, but we estimate a peak value of 50~--~60 (Figure 3). A coronal flare is also detected in the $Chandra$ observations of GJ 699 (Figure 6), acquired approximately 3.5 months (or roughly one stellar rotation period) after the COS observations. The X-ray flare spectra are quantified in the following subsection.


\begin{figure}[h]
\begin{center}
\epsfig{figure=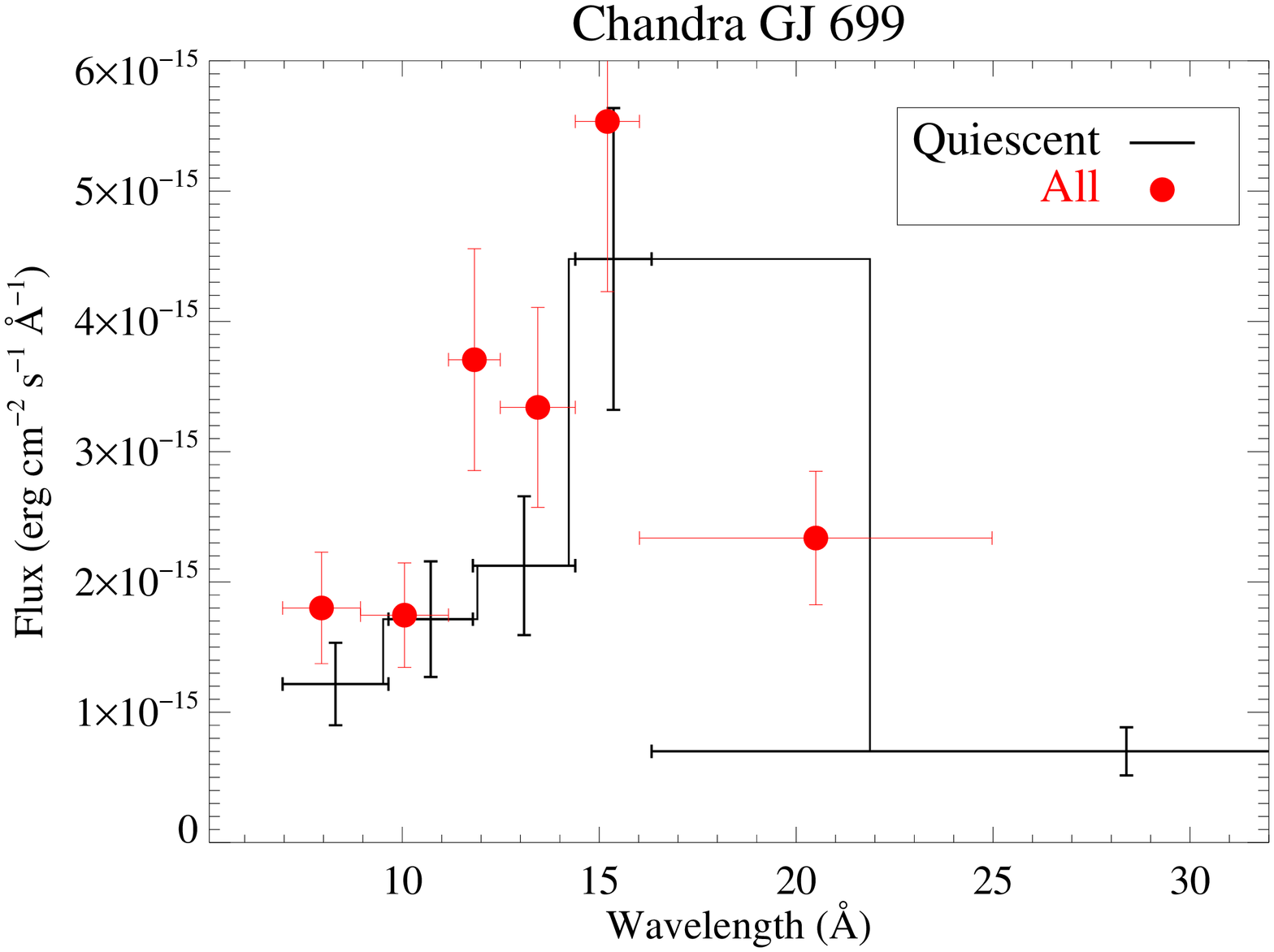,width=2.35in,angle=90}
\epsfig{figure=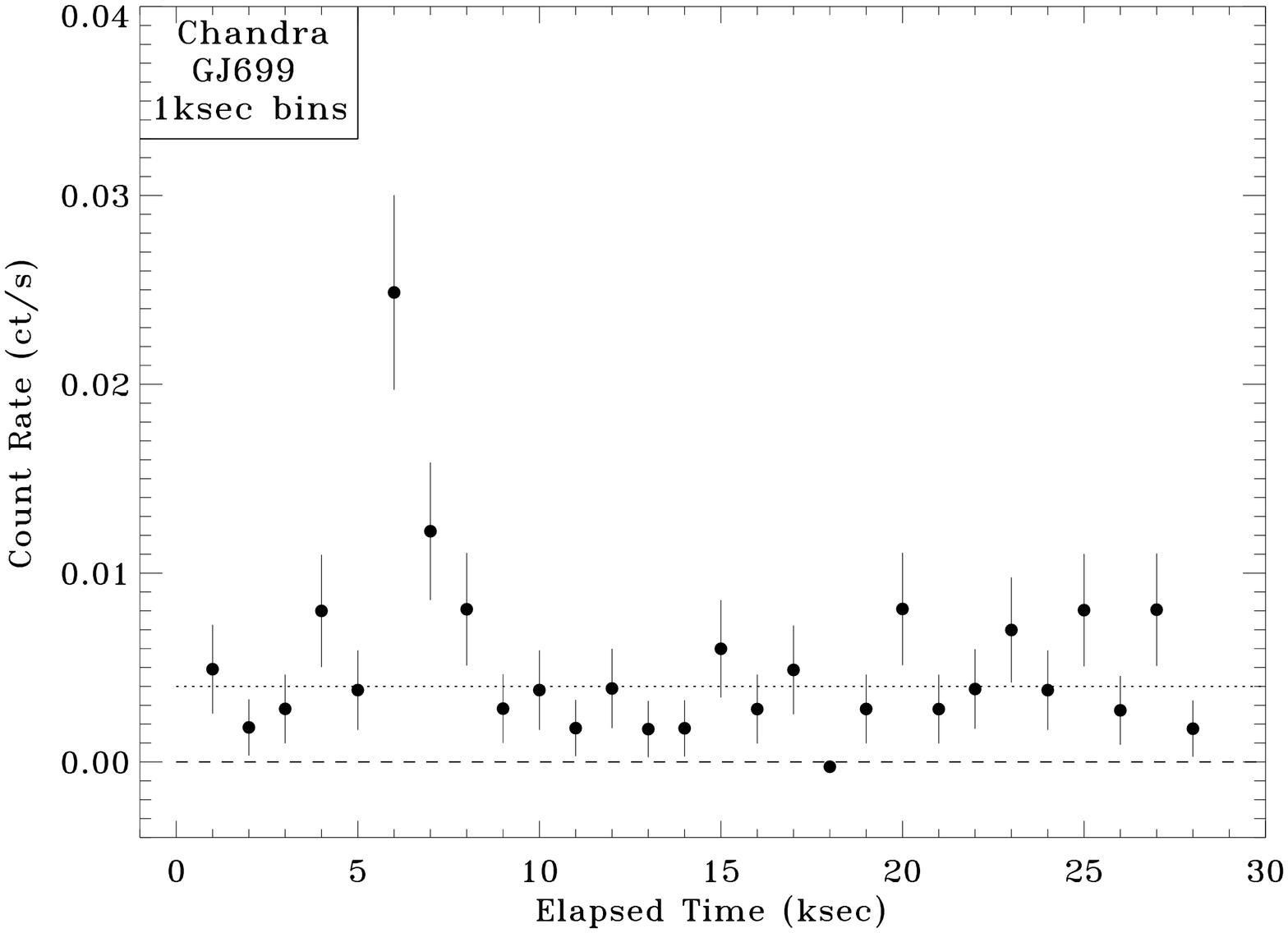,width=2.35in,angle=90}
\vspace{-0.05in}
\caption{
\label{cosovly} $Chandra$ X-ray observations of GJ 699. ($left$): Extracted ACIS spectra, including all observing times (red filled circles) and quiescent times only (black histogram). ($right$): 0.3~--~10 keV (40~$\gtrsim$~$\lambda$~$\gtrsim$~1.25~\AA) lightcurve from the 2019 June 17 visit. One flare is observed starting near 6 ksec. }
\end{center}
\end{figure}

\subsection{X-ray Flares and High-energy Protons from Flare-related CMEs}

We can make order-of-magnitude estimates for the high-energy proton fluxes ($>$ 10 MeV) from coronal mass ejections (CMEs) associated with these large flares by adopting solar X-ray/CME and FUV/CME scaling relations derived by~\citet{cliver12} and~\citet{youngblood17}. We include the following calculations as rough estimates of the particle environment that could be associated with the stellar flares. We note that significant uncertainty exists with the assumption that M dwarf CME behavior is analogous to the Sun given that larger surface magnetic fields on M dwarfs have been hypothesized to confine CMEs to the near-star environment and prevent them from impacting orbiting planets~\citep{alvarado18}.  Furthermore,~\citet{fraschetti19} present simulations of energetic particle trapping in stars with strong magnetic field, with escaping particles concentrated (or, focused) into the orbital plane of the star-planet system.  

{\it Chandra Flare~--~}We measure a peak (0.3 - 10 keV) flare flux  of 2.5~$\times$~10$^{-13}$ erg cm$^{-2}$ s$^{-1}$ and an integrated flare energy of $E_{X}$~~$\approx$~10$^{29.2}$ erg,  with approximately 5.7\% of the total GJ 699 $Chandra$ spectral energy distribution falling in the 1~--~8~\AA\ band. The 1~--~8~\AA\ peak flux in the habitable zone of GJ 699 ($a_{HZ}$~=~0.088 AU, calculated using the analytic formulae of Kopparapu et al. 2014)\nocite{kopparapu14} is estimated to be $F_{X}$(1~--~8\AA)~=~2.6~$\times$~10$^{-4}$ W m$^{-2}$, meaning it would be observed as a roughly X2-3 class equivalent solar X-ray flare, which occur approximately once a month on the Sun. 
The X-ray-to-CME particle flux relationships~\citep{cliver12} describe the frequency and strength of $>$ 10 MeV proton events, when evaluated at 1 AU. As observed at 1 AU, the X-ray flare observed by $Chandra$ is a C-class flare, with a roughly 20\% probability of having an associated CME-like particle outburst~\citep{belov07}. With this caveat in mind, we estimate that this flare could be associated with a peak $>$ 10 MeV proton flux of 0.175 pfu (1 pfu = 1 proton cm$^{-2}$ s$^{-1}$ sr$^{-1}$) at 1 AU from GJ 699, or a HZ proton flux of roughly 23 pfu (typical of an intermediate energy solar CME). 


{\it HST Flares~--~}Analytic relationships between energetic proton fluence and FUV flare flux were developed for M dwarf studies by~\citet{youngblood17}. However, that work focused on relations between \ion{Si}{4} and \ion{He}{2} flares and the proton fluence, not for the suite of spectral lines monitored during the GJ 699 flare observations. \ion{C}{2} follows the flare lightcurve behavior of \ion{Si}{4} almost exactly for a number of large FUV flares~\citep{loyd18a}, with a roughly factor-of-two lower \ion{C}{2}/\ion{Si}{4} ratio during flares compared to the quiescent state~\citep{france16}. We adopt the average \ion{C}{2}/\ion{Si}{4} ratio observed in our STIS G140L observations ($R(C II/Si IV)$ = 2.14), reduced by a factor of two, to estimate the \ion{Si}{4} lightcurve. We find that the \ion{Si}{4} flare fluence, as measured at 1 AU from GJ 699 is 0.0054 J m$^{-2}$. Using Equation 3 from~\citet{youngblood17}, we estimate a peak $>$~10 MeV proton flux of 8.6$^{+16.6}_{-5.6}$ pfu at 1 AU. Scaling this to the average HZ distance of GJ 699, we find peak particle fluxes incident on a HZ planet of approximately 1100 pfu, with factors of $\sim$~3 uncertainty. These values are within the range for what is typical for large (X-class) solar flares accompanied by CMEs~\citep{cliver12}.

As we have shown, both X-ray and FUV observations find relatively energetic flares to be common on GJ 699, despite its advanced age and long rotation period. We observe that high energy photon and particle outputs from GJ 699 are comparable to energetic solar flares. The notable difference is the frequency. Approximately 10\% of the exposure time of the X-ray observations and 40\% exposure time of the FUV observations were dominated by flare emission. A total of 3 X-ray and UV flares in approximately 42 ksec of exposure time is a rate of roughly 6 per day incident on planets orbiting GJ 699.  For comparison, the Sun emitted approximately 4 flares per day (averaged over solar cycle) of GOES classification C, M, and X from 1976~--~2000~\citep{veronig02}, comparable to our estimates.   On the other hand, given the close-in HZ around M dwarfs, C-class solar flares are observed as $\sim$~100~$\times$~higher energy by a planet orbiting GJ 699.  The solar X-class flare frequency alone is much lower, $<$~0.04 flares per day,  approximately 150 times lower frequency than our observed X-ray+UV flare rate on GJ 699.

\begin{figure*}
\begin{center}
\epsfig{figure=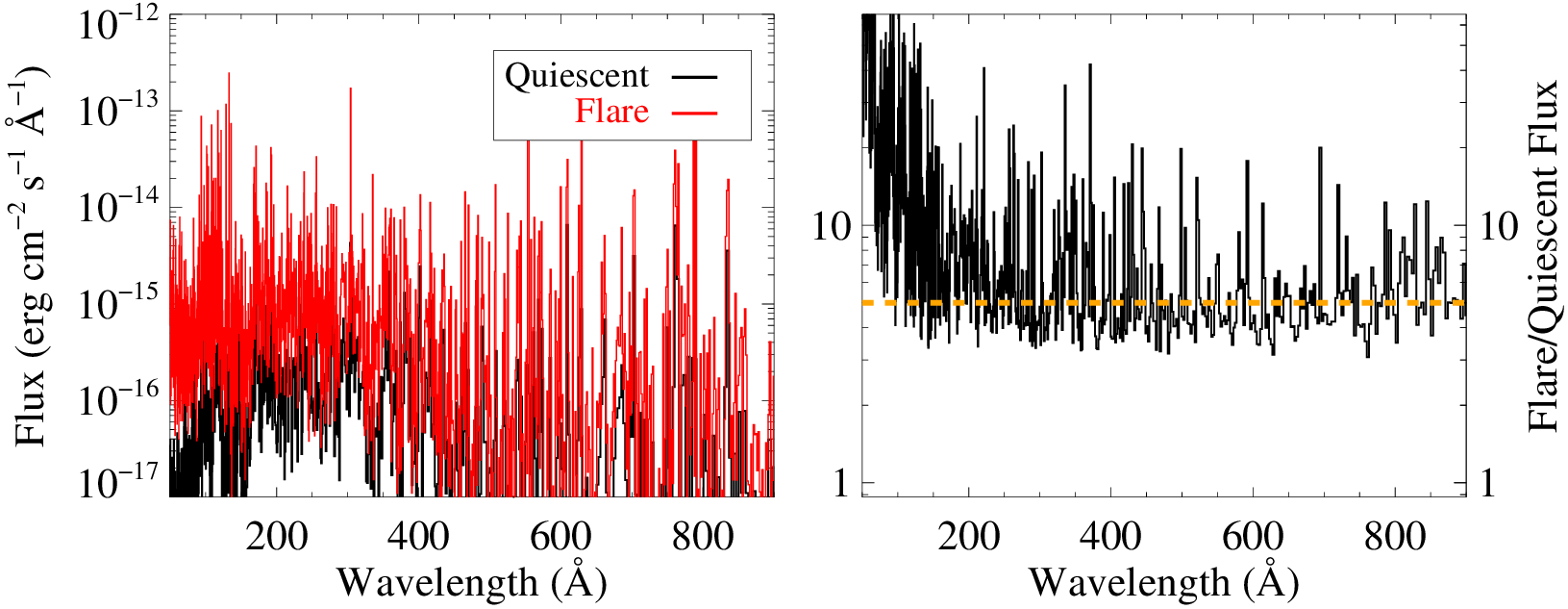,width=3.2in,angle=90}
\vspace{-0.05in}
\caption{
\label{cosovly} EUV spectra of GJ 699 derived from Differential Emission Measure calculations for GJ 699, constructed using observed constraints on the coronal gas (X-rays and FUV iron emission) and chromospheric gas (FUV lines) emission~\citep{duvuuri20}. ($left$): We show the quiescent (black) and flare (red) DEM spectra from 50~--~900~\AA. As expected based on the strong increase in coronal iron lines during the flares observed with $HST$-COS, the flare DEM spectrum is both brighter and increases strongly at higher energies, consistent with EUV flare observations of the active M dwarf AU Mic~\citep{fossi96}. ($right$): The ratio of the flare/quiescent EUV fluxes based on the DEM calculations are plotted to illustrate the predicted changes in the stellar EUV output during flare states. The average flare/quiescent ratio is $\approx$~5 (orange dashed line) with higher ratios observed in lines with higher formation temperatures. We estimate that the flaring EUV spectrum is incident on orbiting planets between 10\% and 40\% of the time based on the observed flare duty cycle. }
\end{center}
\end{figure*}

\subsection{The EUV Spectrum in Quiescent and Flare States}

The Mega-MUSCLES observing program has developed a Differential Emission Measure-based approach to calculating the EUV output of nearby stars~\citep{wilson20}. The DEM calculations used for GJ 699 are based on~\citet{louden17}, and the full description is given by~\citet{duvuuri20}. These calculations use the spectrally-resolved quiescent and flare flux measurements in the FUV and X-ray to constrain the temperature-dependent contribution function and differential emission measure function. The flare and quiescent emission line measurements presented in Table 2, combined with the X-ray spectra shown in Figure 6, span a broad range of formation temperatures, providing constraints on emission formed between 10$^{4}$ $-$ 10$^{5.2}$ K and 10$^{6}$ $-$ 10$^{7}$ K. The differential emission measure is then fit with a 5th order polynomial and interpolated onto the intermediate coronal temperatures that contribute a large fraction of the 100~--~911~\AA\ EUV luminosity. This interpolated differential emission measure function is used to predict the unobserved EUV emission. It is important to note that our DEM approach does not attempt to include continuum emission from atomic ionization edges, e.g., \ion{H}{1} $\lambda$~$<$~912~\AA, \ion{He}{1} $\lambda$~$<$~504~\AA, or \ion{He}{2} $\lambda$~$<$~228~\AA. In the solar spectrum, these continuum sources contribute approximately~20\% to the integrated 100~--~911~\AA\ flux~\citep{woods09}, so our spectra may systematically underestimate the local EUV irradiance seen in the HZ around GJ 699 by a comparable amount. 

Figure 7 ($left$) displays the calculated EUV flux from GJ 699 in both the flare and quiescent states. The right plot in Figure 7 shows the EUV flare/quiescent flux ratio, based on our DEM calculations. The average flux increase during the flare states is observed to be approximately a factor of $\approx$~4~--~5, corresponding to the rough average of the flare increase in the 10$^{4}$~--~10$^{5}$ K emission lines. The band-integrated (100~--~911 \AA) flare spectrum is a factor of 4.4 brighter than the quiescent spectrum of GJ 699. Specific EUV emission lines with higher formation temperatures (e.g., \ion{Fe}{15}~$\lambda$~284~\AA, \ion{Fe}{16}~$\lambda$~335~\AA, etc) have flare/quiescent ratios in the range 20~--~50, reflecting the larger jumps in flare flux observed in the $HST$-COS observations of \ion{Fe}{19} and \ion{Fe}{21}. This is also seen in the larger flare/quiescent ratio at wavelengths shorter than 120~\AA, as these wavelengths are dominated by emission from higher temperature coronal lines.

\subsection{5~\AA~--~10~$\mu$m Spectra in Quiescent and Flare States}

\citet{wilson20} presented the panchromatic spectrum of TRAPPIST-1, constructed from the Mega-MUSCLES program. We adopt the same methodology for GJ 699 to produce complete 5~\AA~--~10~$\mu$m spectral energy distributions for use in modeling planets around old M dwarf stars. These data will be made publicly available through the MUSCLES team High-Level Science Products page at the Mikulski Archive for Space Telescopes\footnote{ {\tt https://archive.stsci.edu/prepds/muscles/}}.

\section{Discussion}

Barnard's Star hosts a cold super-Earth mass planet, but we focus on the star as a proxy to consider the ``typical'' impact on HZ planets around old, mid-M dwarfs. For the purposes of the atmospheric loss calculations presented below, we simply define the ``flare'' spectrum as the coaddition of data obtained in $HST$ orbits 2 and 4 (Flare 1 + Flare 2) plus the X-ray flare periods and the ``quiescent'' spectrum as the coaddition of the data obtained in $HST$ orbits 1, 3, and 5 plus the quiescent X-ray data. While this washes over details about the relative emission line contributions to the different types of flares, presenting an average flare spectrum with a $\sim$~25\% duty cycle provides a general picture of the radiation environment enhancements that orbiting planets could expect to see from GJ 699 at its present age and activity level. In the following two sub-sections, we estimate the thermal and ion escape rates from a hypothetical Earth-like planet orbiting in the HZ of GJ 699, subjected to these high-energy radiation conditions. 

\subsection{Thermal Escape from HZ Planets around Old M Dwarfs}

While non-thermal ion escape dominates atmospheric mass loss from modern Earth, thermal escape, powered by elevated EUV luminosity from the young Sun, may have been the most important escape mechanism on terrestrial planets in the first $\sim$~1 Gyr of solar system history~\citep{tian08, tu15, Amerstorfer17}. Thermal escape in the hydrodynamic regime is thought to dominate the well-studied mass loss from short-period gaseous planets, owing to their high EUV irradiance~\citep{vidal03,ehrenreich15,bourrier16,bourrier18}, and may also dominate  atmospheric escape from rocky planets orbiting active M dwarfs~\citep{luger15,ribas16,bolmont17}. We present a calculation of the atmospheric impacts of GJ 699's high-energy radiation in both quiescence and flare states by placing a hypothetical modern Earth-like planet in the HZ around this star ($a_{HZ}$~=~0.088 AU). In these model calculations, we assumed a rocky planet identical to modern Earth with respect to atmospheric composition, mass, and radius.  The model planet does not include a permanent magnetic field and has a different temperature and pressure structure, driven by the empirically-derived EUV irradiance from GJ 699.  

We developed a parametric model to produce a realistic planetary atmosphere temperature profile by using heating and cooling rates as a function of altitude (following Stevens et al. 1993).\nocite{stevens93}   At the atmospheric levels most strongly affected by EUV heating (and contributing to the subsequent thermal escape; $>$~250 km), we assume that photoionization of oxygen and subsequent electron collisions are the dominant heating term, balanced primarily by downward heat conduction to the mesopause ($\sim$~140km), where radiative cooling becomes significant (see, e.g., Roble et al. 1987). While noting that the higher EUV fluxes from GJ 699 could drive chemical evolution through enhanced molecular dissociation, we do not consider this possibility and hold the mean molecular weight of the model atmosphere constant with pressure.\nocite{roble87}  

We validated our parametric approach by comparing it to Earth atmospheric and solar conditions on March 20, 2014, which approximate mean atmospheric and solar activity conditions.  We employed the MSIS database to acquire an Earth atmospheric temperature and composition profile\footnote{{\tt https://ccmc.gsfc.nasa.gov/modelweb/models/msis\_vitmo.php} } at 60\arcdeg N latitude on the Greenwich meridian, and a solar XUV flux from the TIMED-SEE solar monitoring instrument (1 - 1180\AA, 6.6 mW m$^{-2}$~$\equiv$~F$_{\odot,XUV}$).  Our parametric model atmosphere was required to match the observed exospheric temperature, 1095 K.  The 1095 K exospheric temperature was reproduced with an EUV heating efficiency $\epsilon$ = 0.32.  Thus able to reproduce zeroth-order observables for typical Sun-Earth conditions, we proceeded with using this parameterized atmosphere to explore the influences of the GJ 699 radiation field.  

As a second check on our methodology, we compare our parameterized escape calculations with the more detailed model of~\citet{tian08}, using F$_{*,XUV}$ = 3.8 F$_{\odot,XUV}$ (corresponding to their 4.9 solar flux case\footnote{In this work, we use the TIMED-SEE baselines solar XUV flux of 6.6 mW m$^{-2}$; the Tian et al. baseline solar XUV flux was somewhat lower, 5.1 mW m$^{-2}$}). We compute an exospheric temperature of $\approx$~4800 K, in approximate agreement with their detailed model calculation ($\approx$~4000 K). 
The 1~--~911~\AA\ fluxes for the quiet and flaring spectra of GJ 699 are 1.9 F$_{\odot,XUV}$ and 10 F$_{\odot,XUV}$, respectively. For the quiescent GJ 699 input, we find an exospheric temperature of 2130K, which is marginally higher than the solar maximum dayside temperature at Earth, suggesting comparable atmospheric structure (with an exobase altitude of $\sim$~500 km) as modern day Earth for quiescent illumination conditions.  

However, under constant illumination from GJ 699's flare spectrum, 10 F$_{\odot,XUV}$, the exobase temperature raises to 15,970 K at a distance of 3.3 R$_{\oplus}$.   The important caveat here is that these calculations assume that the flare spectrum illumination occurs in the steady-state.  The atmospheric response time may dampen these effects, depending on the flare duration and duty cycle (see below).   As the atmosphere heats and exobase radius increases, the Jeans thermal escape parameter ($\lambda_{J}$~=~$GMm$/$kT_{exo}$$r_{exo}$; where $T_{exo}$ and $r_{exo}$ are the temperature and radius of the exobase, respectively) decreases and thermalized particles can readily escape the potential well of the planet.  The dimensionless Jeans escape parameters under these conditions for neutral O and H are 2.3 and 0.1 respectively (and lower for ions). This places the atmosphere clearly in the hydrodynamic escape regime.
In this case, the general energy-limited escape formula applies (e.g., Erkaev et al. 2007),\nocite{erkaev07} 
\begin{equation}
\dot{M}~=~\frac{\epsilon \pi F_{*,XUV} r^{2}_{h} R_{\oplus}}{G M_{\oplus}}
\end{equation} 
where the heating efficiency, $\epsilon$ = 0.32 (from the fit to the MSIS exospheric temperature profile) and the `XUV radius' ($r_{h}$~=~1.14~$R_{\oplus}$) is the peak heating altitude in our calculations. Assuming that the EUV heating rate is driven by the flare spectrum times the flare duty cycle ($\sim$~25\%), and that averaging these two states according to the duty cycle is an accurate approximation (see also Bisikalo et al. 2018), the atmospheric mass loss rate is approximately 1.4~$\times$~10$^{7}$ g s$^{-1}$ from a hypothetical Earth-like planet orbiting in the HZ of GJ 699. This indicates that a 1 bar atmosphere (M$_{atm,\oplus}$~=~5.15~$\times$~10$^{21}$ g) is lost in approximately 11 Myr (or nearly 87 Earth atmospheres in a Gyr).\nocite{bisikalo18} 

We note that at EUV illumination levels of 5~--~10 F$_{\odot,XUV}$,~\citet{tian08} find that the combination of hydrodynamic flow and adiabatic cooling can decrease the average exobase temperature and return the atmosphere to hydrostatic equilibrium.   These effects are not considered here; 
a global escape simulation that simultaneously treats these physical effects simultaneously, as well as different compositions, for thermal and non-thermal processes would be valuable. 

\begin{figure*}
\begin{center}
\epsfig{figure=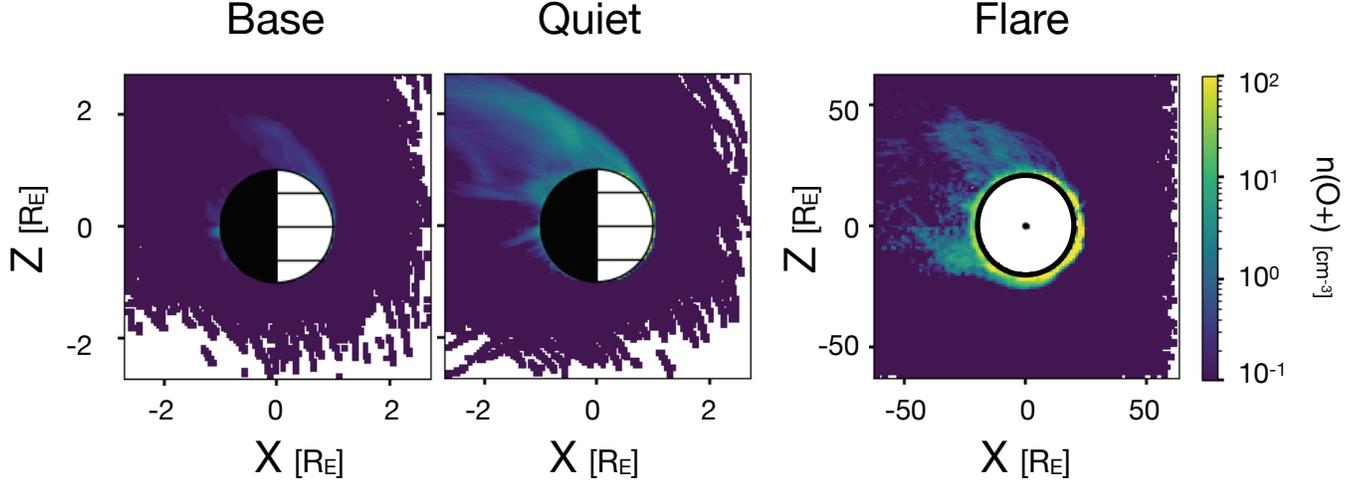,width=7.2in,angle=0}
\vspace{-0.15in}
\caption{
\label{cosovly} O$^{+}$ escape simulations were performed using the hybrid plasma model RHybrid \citep{2018JGRA..123.1678J}. The results are shown for three example cases: 
($left$) unmagnetized Earth orbiting the present-day Sun, ($center$) an unmagnetized, but otherwise Earth-like planet orbiting in the habitable zone ($a_{HZ}$~=~0.088 AU) of Barnard's star in the quiescent stellar condition, and ($right$) the unmagnetized Earth-like planet orbiting in Barnard's star's HZ during stellar flare conditions. The models are displayed in the Planet Stellar Orbital (PSO) coordinate system, such that the planet is at the origin, the x direction points from the planet against the undisturbed upstream stellar wind velocity vector, the z direction is perpendicular to the orbital plane of the planet. Physical distances are shown in planetary radii ($R_p$) that correspond to one Earth radius~--~we note that the distance scale on the Flare model simulation has been increased by a factor of 25 to account for the larger O$^{+}$ extent at the density dynamic range of the Base and Quiet models. We observe that the quiescent XUV spectrum of GJ 699 drives a mild increase in the escape rate relative to modern day Earth, but steady-state illumination by the flare spectrum drives the planet to a much higher atmospheric loss regime.}
\end{center}
\end{figure*}

\subsection{Ion Escape from HZ Planets around Old M Dwarfs}

Ion escape (primarily O$^{+}$) through a polar wind dominates the atmospheric mass loss on Earth today~\citep{seki01}. In order to estimate the atmospheric stability of temperate, terrestrial planets around old, relatively inactive M dwarf stars, we calculated atmospheric oxygen ion escape rates using the hybrid plasma model RHybrid \citep{2018JGRA..123.1678J}, a three-dimensional plasma interaction code where ion macroparticles input from the lower atmosphere (inner boundary in the simulation) and the stellar wind (one side of the outer boundary) interact kinetically with electric and magnetic fields created from the resultant plasma conditions.  Parallelized RHybrid and its sequential predecessor code HYB have been used to study ion escape from Mars \citep{2002JGRA..107.1035K, 2010Icar..206..152K, JGRA:JGRA21894}, Venus \citep{2009AnGeo..27.4333J, 2013JGRA..118.4551J}, Mercury \citep{2003AnGeo..21.2133K}, Titan \citep{2007GeoRL..3424S09K}, and most recently, exoplanets around M-dwarfs \citep{egan_2019}. We adopted the EUV-affected atmospheric thermal results from the preceding subsection, assuming a static atmospheric structure without an adiabatically expanding planetary wind. In light of uncertain exoplanetary magnetic conditions~\citep{lanza15, shkolnik18, cauley19} for short-period, likely tidally-locked, planets around M dwarfs, we elected to study an unmagnetized planet for computational simplicity. Egan and collaborators have demonstrated that the presence of weak planetary magnetic fields can either enhance or inhibit ion escape (for Martian analogs, this inflection point on the magnetic field impact is near 50 nT),  depending on the pressure balance with the stellar wind~\citep{egan_2019}.   We expect that Earth-like magnetic field strengths would cause a net decrease in the ion escape rate.   

Unlike studies of solar system planets, where the ionospheric properties can be validated by direct observations, the boundary conditions of our unmagnetized Earth are estimated from a combination of data derived from Venus, Earth, and Mars. This assumption adds uncertainty about the absolute loss rates predicted by the model, but as a reference point, we expose our model planet to the radiation environment of the modern Earth. This provides a ``base case'' against which the Barnard's star planet models can be compared. Therefore, while the absolute escape rates carry significant uncertainty, the relative rates with respect to an unmagnetized Earth-Sun analog are more robust. 

Figure 8 shows the results of the ion loss calculations, with the O$^{+}$ density shown for the base case, Barnard's star in quiescence, and Barnard's star during a flare period. Ionospheric production was calculated using absorption and ionization rates from \citet{schunk2009ionospheres}. Stellar wind conditions (Table 3) were adopted from the  Barnard's star simulations from~\citet{alvarado19}, extracted at the location of the GJ 699 HZ.  The variability between the different magnetic proxies at the GJ 699 HZ is minor, and we adopted the `Proxy 3' (HD 179949) wind conditions here~\citep{alvarado19}.  Each simulation was run on a $240^3$ grid, 
for 100,000 time steps with $\Delta t = 0.01 s$. 
Figure 8 demonstrates the general behavior of the loss processes studied here: the quiescent XUV spectrum of GJ 699 drives a mild increase in the escape rate relative to modern day Earth, but steady-state illumination by the flare spectrum drives the planet to a much higher atmospheric loss regime. The large increase in the ion escape rate during flare periods (quantified below) is driven by the large increase in the exobase radius compared to the quiescent state ($\propto$~$R_{exo, flare}^{2}$/$R_{exo, quiesc}^{2}$), which increases the cross-section for stellar photon and wind interactions. 

We calculate the O$^{+}$ loss rate in the base case (unmagnetized Earth) to be 2.0~$\times$~10$^{24}$ ions s$^{-1}$.  This number can be compared with the range of observed polar wind ion loss rates observed at Earth today, roughly 10$^{24-26.5}$ ions s$^{-1}$~\citep{seki01,yau88,strangeway05}.
For the hypothetical GJ 699 HZ planet, we find that the ion loss rates are factors of $\sim$~8 and 38,000 times larger than the base case Earth-like planet when GJ 699 is in quiescent (1.6~$\times$~10$^{25}$ ions s$^{-1}$) and flare (7.6~$\times$~10$^{28}$ ions s$^{-1}$) states, respectively.\nocite{seki01,yau88,strangeway05}  We note that in future work it would be interesting to model the thermal and non-thermal escape processes self-consistently to investigate if resupply rates from the lower atmosphere are sufficient to sustain the high O$^{+}$ escape rates.   

Noting the caveats about the absolute ion escape rates, we can estimate the total mass loss rates from a present-day Earth-like planet orbiting Barnard's star. Taking an average flare duty cycle of 25\%, we find total O$^{+}$ loss per Gyr of  (0.75*1.6~$\times$~10$^{25}$ ions s$^{-1}$) + (0.25*7.6~$\times$~10$^{28}$ ions s$^{-1}$)   $\sim$~5.1~$\times$~10$^{5}$ g s$^{-1}$ (1.6~$\times$~10$^{22}$ g Gyr$^{-1}$). Taking the present-day mass of Earth's atmosphere (5.15~$\times$~10$^{21}$ g), the average ion mass loss rate from a HZ planet around Barnard's star is $\sim$~3~Earth atmospheres per billion years. This is comparable to calculations of HZ planets around more active M dwarfs, where complete 1 bar N and O atmospheres are predicted to be lost in less than 0.5 Gyr~\citep{garcia17, airapetian17}. We conclude that HZ planets around old M dwarfs are likely stable to the rapid ion escape in quiescence, but flare activity may drive atmospheric loss rates comparable to their younger cousins orbiting more magnetically exuberant M dwarfs.

The atmospheric calculations presented here are for thermal and ion escape from the specific case of an unmagnetized Earth-analog. We did not consider CME-driven atmospheric changes or stripping~\citep{lammer07,segura10,tilley19} and there are large uncertainties about whether CMEs can escape magnetic confinement on magnetically active M dwarfs to impact their planets~\citep{alvarado18}. We note that atmospheric loss calculations taking into account CMEs would be of particular interest in the case of Barnard's star where surface fields may be low enough for high-energy particles to erupt into the planetary environment~\citep{alvarado19b}. 

Within the framework of these assumptions however, the much higher heating and escape parameters from the flare spectrum drives us towards the conclusion that unless the UV flare duty cycle is very low (e.g., Kowalski et al.  2009; Medina et al. 2020), atmospheric loss will be dominated by the flare periods around old M dwarfs. If the UV flare duty cycle is non-zero, we predict that old M dwarf planets that have acquired second generation atmospheres will quickly lose them to flare-driven EUV heating periods.  GJ 699 is a particularly interesting star in this regard as one of its most notable features is the lack of optical activity and infrequent optical flaring (see, e.g., Paulson et al. 2006; Artigau et al. 2018 and references therein).  GJ 699 stands out as an example of an optically quiet old M dwarf where unexpected levels of high-energy activity may preclude the development and maintenance of a habitable atmosphere.\nocite{artigau18,medina20, kowalski09}   

\subsection{From Atmospheres to Biology on HZ Planets around Old M Dwarfs}

 While a thick atmosphere may be one requirement for a habitable environment, the development of active biology and the resultant biomarkers that are the target of current and future terrestrial planet searches also require the initiation of life on these worlds.  Recent laboratory and theoretical work (e.g.,  Powner et al. 2009; Ritson \& Sutherland 2012; Patel et al. 2015) has shown that NUV photons may drive the formation of RNA nucleotide precursors on temperate, rocky planets with Early-Earth atmospheric compositions~\citep{ranjan17,rimmer18}.\nocite{powner09,ritson12,patel15}   These authors have noted that inactive M dwarfs may have insufficient NUV output to kickoff these prebiotic photochemical chains (see also G{\"u}nther et al. 2020).\nocite{buccino07,guenther20}  Barnard's Star's current day average NUV flux in the habitable zone is $\approx$~6~$\times$~10$^{8}$ photons cm$^{-2}$ s$^{-1}$  \AA$^{-1}$, approximately a factor of 10 below the abiogenesis threshold defined by~\citet{rimmer18}.    
 
 Strong flaring (time-averaged NUV flux enhanced by factors of $\sim$~10) could support prebiotic pathways (see, e.g., Buccino et al. 2007).  On the other hand, while a low UV flare rate may permit stable second generation atmospheres, these worlds may lack sufficient photocatalyzing flux to initiate an active surface biology.   NUV flare rates for M dwarfs in general are not well known, only a small number of stars have time-resolved NUV observations covering the full 2000~--~3000~\AA\ band over sufficient baseline to build up flare statistics~\citep{hawley07, kowalski19}.  Additional NUV flare observations would be valuable for addressing these questions, and we note that the full 2000~--~3000~\AA\ bandpass is important to cover both the emission lines (e.g., \ion{Fe}{2} and \ion{Mg}{2}) and continuum that changes spectral shape during flares~\citep{kowalski19}.   

Finally, if flares are sufficient to prevent the retention of secondary atmospheres in the HZs around old M dwarfs, perhaps the search for atmospheric and biosignature targets will need to be extended further out in the planetary system where planets may retain a large H$_{2}$ envelop and where the XUV environment is less extreme. \citet{pierrehumbert11} proposed an extended habitable zone with liquid water conditions at 1.5 AU around an early M dwarf.   \citet{seager13,seager20} proposed potential biosignatures in H$_{2}$ dominated atmospheres (e.g., nitrous oxide, ammonia, methanethiol, dimethylsulfide, carbonyl sulfide, and isoprene) in extended HZs.\nocite{pierrehumbert11,seager13,seager20}

\section{Summary}

We analyzed new {\it Hubble Space Telescope} and {\it Chandra X-ray Observatory} observations of the $\sim$~10 Gyr mid-M dwarf Barnard's Star to place empirical constraints on the high-energy environment around a representative, old M dwarf. Studies of atmospheric escape suggest that rapid atmospheric loss is likely on terrestrial planets orbiting young M dwarfs, so we were motivated to consider the stability of Earth-like atmospheres later in the main sequence lifetimes of these stars, provided these worlds are able to outgas or capture second-generation atmospheres. The UV and X-ray data, obtained through the Mega-MUSCLES observing program, are used to estimate the currently unobservable EUV radiation that is the major driver of most atmospheric escape processes. 

Over approximately 25 ksec of $HST$ and 27 ksec of $Chandra$ (non-simultaneous) spectroscopy, we observe three intermediate energy flares. We used the flare and quiescent spectral energy distributions to estimate the atmospheric mass loss rates for thermal and non-thermal processes on a hypothetical, unmagnetized terrestrial planet in the habitable zone ($r_{HZ}$~$\sim$~0.1 AU) of Barnard's Star. Our modeling results indicate that the quiescent UV and X-ray flux of Barnard's star elevates the exospheric temperature of our model planet's atmosphere relative to the modern Sun-Earth system, but does not result in catastrophic atmospheric mass loss rates. By contrast, the XUV flare luminosity is large enough to drive significant atmospheric expansion.  
Assuming steady-state illumination of the XUV flare spectrum at the measured UV/X-ray flare duty cycle ($\sim$~25\%), the elevated XUV flux during flare periods dramatically increases the predicted escape rates for all physical processes. We find atmospheric loss of up to $\sim$~90 Earth atmospheres 
per Gyr is possible for our model Earth-like planet.  We add the important caveat that these results are based on a reconstruction of the high-energy spectrum and simplified escape models, uncertainties exist in each step that could alter the overall escape rate and the quantitative conclusions of our study.   While there are significant uncertainties in the absolute level of the total atmospheric escape, the broad conclusion is that if the XUV flare duty cycle is comparable to that observed in typical field age mid-M stars (as it is in this case study of Barnard's Star), then rapid atmospheric mass loss may still be likely on planets around old M dwarfs. This suggests that the XUV flare duty cycle may be one of the most critical stellar parameters when considering the habitability of M dwarf planets.

\acknowledgments
The $HST$ observations presented here were acquired as part of the Cycle 25 Mega-MUSCLES program (15071), supported by NASA through a grant from the Space Telescope Science Institute, which is operated by the Association of Universities for Research in Astronomy, Inc., under NASA contract NAS5-26555. This work was supported by Chandra Guest Observer grant GO8-19017X (ObsID 20619) from Smithsonian Astrophysical Observatory to the University of Colorado at Boulder.  This work made use of the CIAO software package.  KF acknowledges the hospitality of White Sands Missile Range, where a portion of this work was carried out. \\

\bibliography{ms_emapj_gj699}

AUTHOR AFFILIATIONS: \\
(1){Laboratory for Atmospheric and Space Physics, University of Colorado, 600 UCB, Boulder, CO 80309; kevin.france@colorado.edu}\\
(2){Department of Astrophysical and Planetary Science, University of Colorado, 389 UCB, Boulder, CO 80309, USA}\\
(3){Center for Astrophysics and Space Astronomy, University of Colorado, 593 UCB, Boulder, CO 80309}\\
(4){Lunar and Planetary Laboratory, University of Arizona, 1629 East University Boulevard, Tucson, AZ 85721-0092, USA}\\
(5){McDonald Observatory, University of Texas at Austin, Austin, TX 78712, USA}\\
(6){Leibniz Institute for Astrophysics Potsdam, An der Sternwarte 16, 14482 Potsdam, Germany}\\
(7){Harvard-Smithsonian Center for Astrophysics, 60 Garden Street, Cambridge, MA 02138, USA}\\
(8){Institute for Applied Computational Science, Harvard University, 33 Oxford Street, Cambridge, MA 02138, USA}\\
(9){Carl Sagan Institute Cornell University, Space Science Building 311, Ithaca, USA}\\
(10){JILA, University of Colorado and NIST, 440 UCB, Boulder, CO 80309, USA}\\
(11){School of Earth and Space Exploration, Arizona State University, Tempe, AZ 85287, USA}\\
(12){Instituto de Astronom\'ia y F\'isica del Espacio (UBA-CONICET) and Departamento de F\'isica (UBA), CC.67, suc. 28, 1428, Buenos Aires, Argentina}\\
(13){SRON Netherlands Institute for Space Research, Sorbonnelaan 2, 3584 CA, Utrecht, The Netherlands}\\
(14){Leiden Observatory, University of Leiden, Niels Bohrweg 2, 2333CA Leiden, The Netherlands}\\
(15){University Oxford, Atmospheric, Oceanic, and Planetary Physics Department, Clarendon Laboratory, Sherrington Road, Oxford OX1 3PU, UK}\\
(16){Hamburger Sternwarte, University of Hamburg, 21029 Hamburg, Germany}\\
(17){Macau University of Science and Technology}\\
(18){Instituto de Astronom\'ia y F\'isica del Espacio (UBA-CONICET) and UNTREF, CC.67, suc. 28, 1428, Buenos Aires, Argentina}





\begin{deluxetable}{lccc}
\tabletypesize{\normalsize}
\tablecaption{$HST$ and $Chandra$ Observations of Barnard's Star (GJ 699) \label{lya_lines}}
\tablewidth{0pt}
\tablehead{ 
\colhead{Observatory} & \colhead{Mode} & 
\colhead{ Date } & 
\colhead{ Exposure Time (s)} }
\startdata
$Hubble$ & COS G130M & 04 March 2019 & 12,902 \\ 
$Hubble$ & COS G230L & 04 March 2019 & 318 \\
$Hubble$ & STIS G140M & 04 March 2019 & 5,932 \\ 
$Hubble$ & STIS G140L & 04 March 2019 & 7,018 \\ 
$Hubble$ & STIS G230L & 04 March 2019 & 200 \\ 
$Hubble$ & STIS G430L & 04 March 2019 & 5 \\ 
\tableline
$Chandra$ & ACIS-S & 17 June 2019 & 26,700 \\ 
\enddata
\end{deluxetable}

\begin{deluxetable}{lccccc}
\tabletypesize{\small}
\tablecaption{Quiescent and Flare UV Fluxes Observed in $HST$-COS Monitoring Observations\label{lya_lines}}
\tablewidth{0pt}
\tablehead{
\colhead{Emission Line} & \colhead{$\lambda_{rest}$} & \colhead{Flux} & 
\colhead{Flux Uncertainty} & \colhead{FWHM} & \colhead{FWHM Uncertainty} \\ 
& (\AA) & (erg cm$^{-2}$ s$^{-1}$) & (erg cm$^{-2}$ s$^{-1}$) & (km s$^{-1}$) & (km s$^{-1}$) }
\startdata
\tableline
 & & & {\bf Quiescent} & & \\ 
\tableline
C III & 1175.26 & 1.91~$\times$~10$^{-15}$ & 1.54~$\times$~10$^{-16}$ & $-$ & $-$\tablenotemark{a} \\ 
Si III & 1206.50 & 3.60~$\times$~10$^{-16}$ & 1.35~$\times$~10$^{-16}$ & $-$ & $-$\tablenotemark{a} \\ 
N V-a & 1238.82 & 1.36~$\times$~10$^{-15}$ & 0.98~$\times$~10$^{-16}$ & 68.2 & 3.3 \\ 
N V-b & 1242.80 & 6.20~$\times$~10$^{-16}$ & 0.74~$\times$~10$^{-16}$ & 66.8 & 5.7 \\ 
Si II & 1264.74 & 1.34~$\times$~10$^{-16}$ & 0.75~$\times$~10$^{-16}$ & $-$ & $-$\tablenotemark{a} \\ 
C II-a & 1334.53 & 1.05~$\times$~10$^{-15}$ & 0.78~$\times$~10$^{-16}$ & 59.5 & 3.0 \\ 
C II-b & 1335.71 & 1.83~$\times$~10$^{-15}$ & 0.94~$\times$~10$^{-16}$ & 59.0 & 2.1 \\ 
C IV-a\tablenotemark{b} & 1548.20 & 5.27~$\times$~10$^{-15}$ & 6.30~$\times$~10$^{-16}$ & $-$ & $-$\tablenotemark{a} \\ 
C IV-b\tablenotemark{b} & 1550.77 & 2.58~$\times$~10$^{-15}$ & 4.50~$\times$~10$^{-16}$ & $-$ & $-$\tablenotemark{a} \\ 
Mg II-a\tablenotemark{c} & 2795.53 & 6.40~$\times$~10$^{-14}$ & 3.78~$\times$~10$^{-15}$ & $-$ & $-$\tablenotemark{a} \\ 
Mg II-b\tablenotemark{c} & 2802.70 & 4.91~$\times$~10$^{-14}$ & 3.52~$\times$~10$^{-15}$ & $-$ & $-$\tablenotemark{a} \\ 
& & & & & \\ 
Fe XIX & 1118.06 & $<$ 1.0~$\times$~10$^{-16}$ & $\cdots$ & $\cdots$ & $\cdots$ \\ 
Fe XII & 1242.00 & $<$ 1.0~$\times$~10$^{-16}$ & $\cdots$ & $\cdots$ & $\cdots$ \\ 
Fe XXI & 1354.08 & $<$ 0.8~$\times$~10$^{-17}$ & $\cdots$ & $\cdots$ & $\cdots$ \\ 
& & & & & \\ 
H$_{2}$ (1~--~3) Q(3) & 1119.08 & 2.53~$\times$~10$^{-16}$ & 1.00~$\times$~10$^{-16}$ & 17.9 & 4.8 \\ 
H$_{2}$ (1~--~4) Q(3) & 1163.81 & 5.10~$\times$~10$^{-16}$ & 0.80~$\times$~10$^{-16}$ & 31.3 & 3.6 \\ 
\tableline
 & & & {\bf Flare}  & & \\ 
\tableline
C III & 1175.26 & 8.72~$\times$~10$^{-15}$ & 2.54~$\times$~10$^{-16}$ & $-$ & $-$\tablenotemark{a} \\ 
Si III & 1206.50 & 6.50~$\times$~10$^{-15}$ & 8.00~$\times$~10$^{-16}$ & $-$ & $-$\tablenotemark{a} \\ 
N V-a & 1238.82 & 4.54~$\times$~10$^{-15}$ & 2.54~$\times$~10$^{-16}$ & 46.8 & 1.5 \\ 
N V-b & 1242.80 & 2.19~$\times$~10$^{-15}$ & 1.78~$\times$~10$^{-16}$ & 46.9 & 2.5 \\ 
Si II & 1264.74 & 3.60~$\times$~10$^{-16}$ & 2.00~$\times$~10$^{-16}$ & $-$ & $-$\tablenotemark{a} \\ 
C II-a ($narrow$\tablenotemark{d}) & 1334.53 & 3.22~$\times$~10$^{-15}$ & 3.89~$\times$~10$^{-16}$ & 38.0 & 3.5 \\ 
C II-a ($broad$\tablenotemark{d}) & 1334.53 & 1.27~$\times$~10$^{-15}$ & 5.06~$\times$~10$^{-16}$ & 104.6 & 17.2 \\ 
C II-b ($narrow$\tablenotemark{d}) & 1335.71 & 5.98~$\times$~10$^{-15}$ & 5.87~$\times$~10$^{-16}$ & 44.4 & 3.1 \\ 
C II-b ($broad$\tablenotemark{d}) & 1335.71 & 1.65~$\times$~10$^{-15}$ & 6.42~$\times$~10$^{-16}$ & 87.0 & 8.9 \\ 

& & & & & \\ 
Fe XIX & 1118.06 & 6.26~$\times$~10$^{-16}$ & 0.98~$\times$~10$^{-16}$ & 71.5 & 7.6 \\ 
Fe XII & 1242.00 & 1.40~$\times$~10$^{-16}$ & 0.56~$\times$~10$^{-16}$ & $-$ & $-$\tablenotemark{a} \\ 
Fe XXI & 1354.08 & 8.70~$\times$~10$^{-16}$ & 1.10~$\times$~10$^{-16}$ & 115.6 & 11.3 \\ 
& & & & & \\ 
H$_{2}$ (1~--~3) Q(3) & 1119.08 & 5.94~$\times$~10$^{-16}$ & 1.29~$\times$~10$^{-16}$ & 26.4 & 4.3 \\ 
H$_{2}$ (1~--~4) Q(3) & 1163.81 & 7.87~$\times$~10$^{-16}$ & 1.10~$\times$~10$^{-16}$ & 29.0 & 3.1 \\ 
\enddata
\tablenotetext{a}{Indicates line-widths could not be cleanly measured owing to low-S/N, the line appearing at the edge of the $HST$-COS detector segment (Si III), blending (C III), or the line was observed with a low-resolution mode (C IV, Mg II). The C III 1175 multiplet consists of 5 strong components. The flux of C III was integrated over the full extent of the line, but the close line separation prevents robust line-width determination. } 
\tablenotetext{b}{Measured with STIS G140L.} 
\tablenotetext{c}{Measured with STIS G230L.} 
\tablenotetext{d}{During the GJ 699 flare, the C II emission lines clearly displayed a two-component morphology, which we fitted with the superposition of two Gaussians. The broad C II components were observed redshifted by 8.5 and 19.4 km s$^{-1}$ for the 1334 and 1335 components, respectively. } 
\end{deluxetable}

\begin{deluxetable}{lcc}
\tabletypesize{\normalsize}
\tablecaption{Stellar Wind Parameters Assumed for GJ 699\tablenotemark{a} \label{lya_lines}}
\tablewidth{0pt}
\tablehead{ 
\colhead{Parameter} & \colhead{Earth Baseline} & 
\colhead{GJ 699} }
\startdata
B-field Flux [nT] & [3.53, 3.53,0] & [8.57, 8.57, 0] \\ 
speed [km s$^{-1}$] & 400 & 409 \\
density [cm$^{-3}$] & 8 & 200 \\ 
temperature [K] & 10$^{5.1}$ & 10$^{6.2}$ \\ 
\tableline
\enddata
\tablenotetext{a}{Evaluated at 0.088 AU, based on the `Proxy 3' (HD 179949) Barnard's Star stellar wind calculations of \citet{alvarado19}. } 
\end{deluxetable}

\end{document}